\documentclass[fleqn,12pt]{article}
\usepackage{amsfonts,epsfig}
\usepackage[nospace]{cite}

\newcommand{\ft}[2]{{\textstyle\frac{#1}{#2}}}
\def\tilde{\widetilde}
 
\def\1bar{1\hskip -.275cm -}
\def\2bar{2\hskip -.275cm -}
\def\3bar{3\hskip -.275cm -}

\newsavebox{\uuunit}

\makeatletter \@addtoreset{equation}{section} \makeatother

\def\bfone{\relax{\rm 1\kern-.35em 1}}

\def\bfone{\relax{\rm 1\kern-.35em 1}}
 
\newcommand{\nc}{\newcommand}

\newcommand{\CA}{{\cal A}}
\newcommand{\CB}{{\cal B}}
\newcommand{\CC}{{\cal C}}

\newcommand{\Bj}{{\bar{\jmath}}}

\newcommand{\la}{\label}

\newcommand{\Ref}[1]{(\ref{#1})}

\newcommand{\pls}{\!+\!}

\newcommand{\mis}{\!-\!}

\newcommand{\mathon}{\mathversion{bold}}
\newcommand{\mathoff}{\mathversion{normal}}
\newcommand{\ie}{{\it i.e.}}
\newcommand{\eg}{{\it e.g.}}
\newcommand{\ap}{\alpha^{\prime}}

\nc{\be}{\begin{equation}} \nc{\ee}{\end{equation}}
\nc{\bea}{\begin{eqnarray}} \nc{\eea}{\end{eqnarray}}
\newcommand{\ben}{\begin{displaymath}}
\newcommand{\een}{\end{displaymath}}
\nc{\dalpha}{\dot{\alpha}} \nc{\dbeta}{\dot{\beta}}
\nc{\nn}{\nonumber} \nc{\non}{\nonumber\\}
\nc{\snabla}{\nabla\hspace{-3 mm}\slash}
\nc{\spartial}{\partial\hspace{-2 mm}\slash}
\begin{document}

\thispagestyle{empty}

\renewcommand{\thefootnote}{\fnsymbol{footnote}}

\begin{center}
ROM2F/2003/08\hspace*{1.5cm}
ITPUU-03/22\hspace*{1.5cm}
SPIN-03/12
\end{center}

\bigskip

\begin{center}
\mathon
{\bf\Large On stringy AdS$_5\times S^5$ and  higher
spin holography}
\mathoff
\bigskip\bigskip

{\bf Massimo Bianchi\footnote{\tt Massimo.Bianchi@roma2.infn.it}, 
\addtocounter{footnote}{1}
Jose F.~Morales\footnote{\tt  Francisco.Morales@lnf.infn.it},
and Henning~Samtleben\footnote{\tt H.Samtleben@phys.uu.nl}}

\addtocounter{footnote}{-3}
\vspace{.3cm}  
$^\fnsymbol{footnote}$ {\it 
Dipartimento di Fisica and INFN \\
Universit\`a di Roma ``Tor Vergata''\\
00133 Rome, Italy}

\addtocounter{footnote}{2}
\vspace{.3cm}

$^\fnsymbol{footnote}$  {\it 
•Laboratori Nazionali di Frascati,\\
Via E. Fermi, 40\\ I-00044 Frascati (Rome), Italy}

\addtocounter{footnote}{1}
\vspace{.3cm}

$^\fnsymbol{footnote}$  
{\it Institute for Theoretical Physics and Spinoza Institute\\
Utrecht University, Postbus 80.195\\
3508 TD Utrecht, The Netherlands}

\end{center}
\renewcommand{\thefootnote}{\arabic{footnote}}
\setcounter{footnote}{0}
\bigskip
\medskip

\begin{abstract}

We derive the spectrum of Kaluza-Klein descendants of string
excitations on AdS$_5\times S^5$. String states are organized in long
multiplets of the AdS supergroup $SU(2,2|4)$ with a rich
pattern of shortenings at the higher spin enhancement point
$\lambda=0$.  The string states holographically dual to the higher
spin currents of SYM theory in the strict zero coupling limit are
identified together with the corresponding Goldstone particles
responsible for the Higgsing of the higher spin symmetry at
$\lambda\neq 0$. Exploiting higher spin symmetry we propose a very
simple yet effective mass formula and establish a one-to-one
correspondence between the complete spectrum of $\Delta_{0} \le 4$
string states and relevant/marginal single-trace deformations in
${\cal N}=4$ SYM theory at large $N$.
To this end, we describe how to efficiently enumerate scaling
operators in `free' YM theory, with the inclusion of fermionic
`letters', by resorting to Polya theory.  Comparison between the
spectra of $\ft14$-BPS states is also presented.  Finally,
we discuss how to organize the spectrum of ${\cal N}=4$ SYM theory in
$SU(2,2|4)$ supermultiplets by means of some kind of
`Eratostenes's sieve'.

\end{abstract}

\renewcommand{\thefootnote}{\arabic{footnote}}
\vfill

\bigskip


\setcounter{footnote}{0}
\newpage

\tableofcontents

\section{Introduction and Summary}
\label{introsum}

The quest for a string description of gauge theories has a long
history starting from the pioneering work of A.~Polyakov \cite{polwl}
and G.~'t~Hooft \cite{thooN}.  These ideas have recently found a
concrete realization in the remarkable proposal by
J.~Maldacena~\cite{jm} for an exact correspondence between type IIB
superstring on AdS$_{5}\times S^{5}$ with $N$ RR 5-form fluxes and
${\cal N}=4$ supersymmetric Yang-Mills (SYM) theory with $N$ colors.
At large $N$ and fixed $\lambda = g_{_{\rm YM}}^2 N$, non-planar
diagrams corresponding to string loops are suppressed but superstring
quantization in the presence of RR backgrounds is poorly understood
\cite{adsquant,berko}, except possibly for pp-waves \cite{ppwaves,bmn}
that describe the so-called BMN limit \cite{bmn}. For this reason one
has to resort to an effective low-energy supergravity description,
valid at weak curvature (large radius) $R\gg\sqrt{\alpha^\prime}$, that only captures
the strong coupling regime $\lambda \gg 1$ of the gauge theory . As $R$
is reduced towards its minimal value, $R\approx \sqrt{\alpha^\prime}$,
the supergravity description breaks down and stringy physics becomes
relevant. On the SYM side, the theory is driven to its perturbative
regime $\lambda \ll 1$, ending at the highly symmetric point of
vanishing coupling.  The dynamics at this point is highly constrained
if not frozen by the presence of infinitely many higher spin
symmetries, \ie~by an infinite set of conserved currents of
arbitrarily high spins.  Pushing Maldacena duality to its limit,
one may expect a duality to a bulk theory with an infinite number of
massless higher spin (HS) particles on AdS.\footnote{By masslessness
in AdS we understand particles associated to gauge invariances.}

In flat spacetime a non-trivial dynamics in the analogous zero
tension limit of superstring theory seems to be ruled out by the
theorem of S.~Coleman and J.~Mandula. The assumptions of this theorem
are violated by the presence of a non-trivial cosmological
constant, and one may ask whether there can be an interacting
theory of higher spins (HS) on AdS. The answer seems to be
affirmative. The interactions that correspond to joining and
breaking of strings should presumably be governed by $g_s \approx 1/N$
in this limit~\cite{sezsun5}.

This seems exactly what the doctor ordered since progress in
constructing consistent theories for interacting fields of higher spin
has for long been hampered by the assumption that flat spacetime would
be a good starting point. Quite amazingly, on the contrary the best
starting point seems to be AdS \cite{sundb,mikha}. In a remarkable
series of papers M.~Vasiliev and collaborators have developed a
formalism that consistently describes interacting HS theories in
$D=4$~\cite{vas4} and made some progress towards solving the same
problem in $D=5$~\cite{vas5} or even in arbitrary dimension for
totally symmetric tensors \cite{vasD}. A hint on how the former case
can find application comes from the recent proposal of I.~Klebanov and
A.~Polyakov~\cite{klepol} that has been generalized by the work of
several groups~\cite{sezsun4,pet,gir,dasjev,surya}. Free geometric equations for
massless HS gauge fields in flat $D=4$ spacetime have also been
recently proposed \cite{frasag}.

In the search for consistent HS theories in $D=5$, E.~Sezgin and
P.~Sundell~\cite{sezsun5} have shown that superstring states belonging
to the first Regge trajectory on AdS can be put in one to one
correspondence with the physical states in the master fields of
Vasiliev's theory that in turn encompass bilinear (twist two)
composites on the boundary. The linearized bulk field equations assume
the desired form. In particular, one finds the correct self-duality
constraints emerging from the natural curvature constraints. The
energy-spin relation turns out to be linear, $ MR \approx s $, rather
than quadratic $M^2 = (s-1)/\alpha^\prime$ as for
superstrings in flat spacetime.  This linear behavior has been
confirmed by studies of `long' solitonic strings in AdS, as opposed to
`short' ones that behave as in flat space \cite{strisol}. In the
spirit of holography the massless higher spin gauge theory describes
the dynamics of twist two operators with conformal dimensions
saturating the AdS unitarity bound:
\be \Delta ~=~ s + 2 \;.
\ee
Another very interesting aspect of the problem is what happens when
the string tension is small but non-vanishing. Out of the infinite
number of conserved HS currents only a handful are preserved, while the
infinitely many remaining ones are violated. In conformal field theory, 
and this is
the case if one simply turns on the gauge coupling rather than
relevant operators, violating the conservation of a current of any
spin is a symptom of its acquiring an anomalous dimension $\Delta = 2
+ s + \gamma$
\cite{konan,bkrslog,bkrsandim,bkrskon,bersmix,appss}. Global
symmetries on the boundary correspond to local symmetries in the bulk;
thus the holographic counterpart of this is a unprecedented Higgs
mechanism in which infinitely many massless HS particles eat
infinitely many lower spin Goldstone particles and become
massive. Ideally one would like to describe this ``grande bouffe'' at
the fully non-linear level, in the hope that the coupling of the
Goldstone master fields to Vasiliev's master fields, as well as the
self-interaction of the latter, be fixed by higher spin symmetry. Less
ambitiously one may ask whether the resulting `mass shifts' to lowest
order match the one-loop anomalous dimensions that have a strikingly
simple form \cite{dolosb,qcdetal}
\be \gamma^{(s)}_{1-{\rm loop}}(\lambda) ~=~
{\lambda\over 2\pi^2} \sum_{k=1}^{s-2} {1\over k} \;
\ee
where $s$ is the top spin in the (long) multiplet.
In order to do that one may not forgo understanding the structure of
the Goldstone master fields and their possible mixing with an infinite
tower of Kaluza-Klein (KK) master fields that unavoidably appear when the
HS theory is viewed as a consistent truncation of type IIB superstring
compactification on $S^5$.

The aim of the present paper is to set the basis for a systematic
study of these issues. We will focus on the spectrum of
the theory, leaving the issue of interactions to future
investigations. KK reductions of
supergravities on AdS$_p\times S^q$ spaces have been studied in
various contexts \cite{kk,deboer,ms}. Here we apply these techniques 
in order to determine the spectrum of KK descendants of
higher string excitations on AdS$_5\times S^5$.

Before entering the technical description of our results it is worth
spending some words to explain our general philosophy.  There are two
choices for maximally supersymmetric vacua of string theory in
$D=5$. They correspond to compactifications of type IIB theory on
$T^5$ and on $S^5$.  At low energies, the former leads to ${\cal N}=8$
`rigid' (or rather `ungauged') supergravity, while the latter is believed to
be governed by ${\cal N}=8$ $SO(6)$ gauged supergravity. The two vacua
can be `continuously' related by formally turning on and off the gauge
coupling constant in $D=5$.  In both cases, it should be consistent at
tree level to neglect KK descendants as well as various $p$-brane
wrappings and only retain fundamental string
excitations.\footnote{Despite our limited understanding of superstring
quantization on AdS$_5\times S^5$ \cite{adsquant,berko}, the absence
of non-trivial cycles in $S^5$ should be taken as a positive 
indication in
this respect.}  Agreement between the two pictures then requires that
the ground `floor' in the towers of KK descendants of type IIB string
excitations on AdS$_5\times S^5$ accommodate the spectrum of the rigid
theory when the latter is properly rearranged in representations of
the AdS supergroup.  This will be the starting point (or assumption)
of our KK algorithm.

More precisely, to each string excitation on flat space we will associate a
tower of KK descendants on $S^5$.  Proceeding in this way one
automatically ensures that the ground floors in the KK towers on $S^5$
account for the right number of superstring degrees of freedom in the
rigid limit.  The resulting theory can then be thought of as an
$SO(6)$ gauging of type IIB superstring theory dimensionally 
reduced on $T^5$ much in the
same way as ${\cal N}=8$ gauged supergravity can be realized as a
gauging of type IIB supergravity reduced on $T^5$.

Field equations for string fluctuations around AdS vacua are not
available even at the linearized level except for `massless states'
\cite{sezsun5}.  As a result the dynamical information about scaling
dimensions $\Delta$ will be missing in our initial analysis.  Unlike
the remaining $SO(4)\times SO(6)$ quantum numbers in the bosonic
symmetry group that will be unambiguously determined by our KK
algorithm, conformal dimensions are sensible to quantum
corrections. In the planar limit, the spectrum of $\Delta =
\Delta(\lambda)$ should be determined by consistent quantization of
the superstring on AdS$_5\times S^5$~\cite{adsquant,berko}.  Despite
some recent progress there seems to be a long way to go.  Still,
supersymmetry and higher spin symmetries can be exploited in order to
get some insight into the general form of the spectrum of conformal
dimensions $\Delta_{0}$ at $\lambda=0$.  The recent observations in
\cite{bits,lind,lindsun,zeroten,lindzab} raise some hope of a possible direct evaluation
of the spectrum in an effective worldsheet theory.  We hope to explore
these ideas in the near future.

The spectrum of KK descendants that we have derived provides us with a
rather rich set of data on which general ideas about holography beyond
the supergravity approximation can be quantitatively tested. Indeed as
we shall see, the resulting spectrum captures most of the interesting
features of HS AdS/CFT holography. In particular, the gauge multiplets
realizing the expected $hs(2,2|4)$ higher spin symmetry appearing in
${\cal N}=4$ SYM at $\lambda=0$ \cite{ans,sundb,sezsun5} are
identified in the string picture together with their associated
Goldstone particles. Quite remarkably if not unexpectedly, given our
rather bold assumptions, we will establish a one-to-one correspondence
between the spectrum of $\Delta_0\le4$ superstring excitations and
that of relevant and marginal single-trace primary operators of ${\cal
N}=4$ SYM.\footnote{The agreement seems to persist up to scaling
dimension $\Delta<6$, including $\Delta= 5.5$, where a fermionic
superconformal primary first appears. We thank N. Beisert for sharing
his insights on this point with us.} Conformal descendants are
accounted for by the zero-modes of the string coordinates (4d
`momentum').  Mixing with multi-particle states corresponding to
multi-trace operators is suppressed at large $N$
\cite{bersmix,appss,bks} and will not be discussed any further. Nor shall
we discuss the spectrum from the viewpoint of the light-cone
gauge quantization~\cite{metsa}.

The plan of the paper is as follows: In Section~\ref{IIBflat} we
review the construction of the massive string spectrum of type IIB
theory in flat spacetime. In Section~\ref{IIBads} we discuss our
`naive' KK reduction on AdS$_5\times S^5$ and exploit HS symmetry at
small radius.  In Section~\ref{compa} we compare the resulting
spectrum with that of the free ${\cal N}=4$ SYM theory. To this end we
compute in Section~\ref{gauge} the partition function of
gauge-invariant (on-shell) single-trace operators using Polya theory
\cite{polya} and generalizing A.~Polyakov's results \cite{polpol}, by
the inclusion of fermionic ``letters". This section is self-contained
and can be read independently of the rest of the paper. We also
discuss superconformal transformations that are needed to identify
`HWS' (superconformal primaries) together with an alternative 
procedure based on `Eratostene's (super)sieve'. In Section~\ref{conclusions} 
we draw some conclusions. Finally, in two appendices we collect some background
material, a self-contained description of multiplet shortenings,
largely based on the work of F.~Dolan and H.~Osborn \cite{dolosb}, and
several useful tables.

\section{Type II superstrings in flat space }
\label{IIBflat}

We start by reviewing the spectrum of type II superstrings
in flat space \cite{gsw} in the GS formulation wherein 
chiral string excitations are created by the
raising modes  $\alpha_{-n}^I,S^{a}_{-n}$ acting on the
vacuum $|{\cal Q}_s\rangle$:
\bea
 S^{a}_{n}|{\cal Q}_c\rangle &=&
\alpha^I_n|{\cal Q}_c\rangle=0  \quad\quad\quad n>0 \;.
\eea
Here and below indices $I=1, \dots, {\bf 8_v}$, $a=1, \dots, {\bf
8_s}$, $\dot{a}=1, \dots, {\bf 8_c}$ run over the vector, spinor left
and spinor right representations of the $SO(8)$ little Lorentz
group. In addition we have introduced the compact notations:
\be {\cal Q}_s ={\bf 8_v}+{\bf
8_s}
\quad\quad {\cal Q}_c ={\bf 8_v}+{\bf 8_c}
\;,
\label{8vs} \ee 
to describe chiral worldsheet supermultiplets.
The vacuum $|{\cal Q}_c\rangle$ is $2^4$-fold degenerated as a result of
the quantization of the eight fermionic zero modes $S^{a}_{0}$. 
For the first few string excitation levels one finds:
\bea
{\ell=0}\quad &&|{\cal Q}_c\rangle \;, \nn\\
{\ell=1}\quad && (S^{a}_{-1}+\alpha^I_{-1})
 |{\cal Q}_c\rangle \;, \nn\\
{\ell=2}\quad && (S^{a}_{-2}+\alpha^I_{-2}) |{\cal Q}_c\rangle,
\quad S^{[a}_{-1} S^{b]}_{-1} |{\cal Q}_c\rangle ,\quad
\alpha^{(I}_{-1}\alpha^{J)}_{-1} |{\cal Q}_c\rangle,\quad
 S^{a}_{-1}\alpha^I_{-1}|{\cal Q}_c\rangle\;, \nn\\
{\ell=3}\quad && (S^{a}_{-3}+\alpha^I_{-3}) |{\cal Q}_c\rangle,
\quad
(S^{a}_{-2}+\alpha^I_{-2})(S^{a}_{-1}+\alpha^I_{-1})|{\cal Q}_c\rangle
,\quad
 S^{[a_1}_{-1}\dots S^{a_3]}_{-1} |{\cal Q}_c\rangle\;, \nn\\
&&  \quad \alpha^{(I_1}_{-1}\dots \alpha^{I_3)}_{-1} |{\cal Q}_c\rangle \quad
S^{[a_1}_{-1}S^{a_2]}_{-1} \alpha^{I}_{-1}|{\cal Q}_c\rangle
,\quad \alpha^{(I_1}_{-1}\alpha^{I_2)}_{-1} S^{a}_{-1}
|{\cal Q}_c\rangle \;.
 \eea
The physical spectrum of the type IIB  superstring is defined by
tensoring two (left and right moving) identical chiral spectra
subject to the level matching condition $N_L=N_R$. Type IIA string
states are given instead by tensoring chiral spectra with opposite
chiralities.  Decomposing into $SO(8)$ representations 
the  spectrum $T_\ell$ at 
string excitation level $\ell$ yields:
\bea T^{IIA,B}_0 &=& {\cal Q}_c\, Q_{c,s} \;,
\nn\\
T_1 &=& {\cal Q}_c^2 {\cal Q}_s^2  \;,\nn\\
T_2 &=& {\cal Q}_c^2({\cal Q}_s+{\cal Q}_s\cdot {\cal Q}_s )^2 =
T_1\times ({\bf 1}+{\bf 8_v})^2  \;, \nn\\
 T_3 &=& {\cal Q}_c^2({\cal Q}_s+{\cal Q}_s^2+{\cal Q}_s\cdot {\cal Q}_s\cdot 
 {\cal Q}_s)^2=
  T_1 \times ({\bf 1}+{\bf 8_v}+{\bf 35_v}+{\bf 8_s}+{\bf 8_c})^2
\;,
\label{flat}
\eea
and so on. In the right hand sides of (\ref{flat}) we have used
$SO(8)$ group theory in such a way as to factor out $T_1$ that represents 
the contribution of a 'minimal' massive multiplet with exactly $2^{16}$ states. 
Thanks to the presence of the overall ${\cal Q}_c^2$ we need only factor out 
two ${\cal Q}_s$, each coming from rewriting dotted (graded symmetrized) 
products according to
\bea
{\cal Q}_s\cdot
{\cal Q}_s &\equiv & 8_{(v}\times 8_{v)}+8_{[s}\times 8_{s]}+8_v\times
8_s={\bf 8_v}\times {\cal Q}_s \;,\nn\\
{\cal Q}_s\cdot {\cal Q}_s \cdot {\cal Q}_s &\equiv & 8_{(v}\times 8_v\times
8_{v)}+8_{[s}\times 8_s\times  8_{s]}+8_v\times 8_{[s}\times
8_{s]}+8_s\times 8_{(v}\times 8_{v)} \nn\\
&=& ({\bf 35}+{\bf 8_c})\times {\cal Q}_s \;. \label{fcs}
 \eea
Clearly the spectrum of massive superstring excitations being non-chiral
makes no distinction between type IIA and type IIB theories. This can be 
seen in (\ref{flat}) where beyond $T_0$, string
excitations always come in left-right symmetric representations of
$SO(8)$. Consequently, the spectrum of $\ell\geq 1$ string excitations
in (\ref{flat}) can be reorganized in terms of the $SO(9)$ little
Lorentz group of a massive particle in ten dimensions
\bea
T_1 &=& \left([2,0,0,0]+[0,0,1,0]+[1,0,0,1]\right)^2 ~=~
\left({\bf 44+84+128}\right)^2 \;, \nn\\
T_\ell &=& T_1 \times ({\rm vac}_\ell)^2 \;,
\label{tn}
\eea
with
\bea
{\rm vac}_1 &=& [0, 0, 0, 0] ~=~ {\bf 1} \;,
\nn\\
{\rm vac}_2 &=& [1, 0, 0, 0] ~=~ {\bf 9} \;,
\nn\\
{\rm vac}_3 &=& [2, 0, 0, 0]+[0, 0, 0, 1] ~=~
     {\bf 44}+{\bf 16} \;,  \nn\\
{\rm vac}_\ell &=& [\ell-1, 0, 0, 0]+\ldots \;.
\eea
where $[a,b,c,d]$ are $SO(9)$ Dynkin labels.
Notice that fermionic HWS, corresponding to ${\bf 44}\times{\bf 
16}$ and ${\bf 16}\times{\bf 
44}$, first appear at level $\ell = 3$.
As an illustration, we have listed the ${\rm vac}_\ell$ for the lowest
massive levels $\ell\le8$ in appendix~\ref{app:vac}.

\section{Type IIB string on AdS$_5\times S^5$ }
\label{IIBads}

To linear order in fluctuations around AdS$_5\times S^5$ and after 
diagonalization, the type IIB
field equations should boil down to a set of uncoupled free
massive equations: 
\bea &&\left(\nabla_{AdS_5\times S^5}^2-M_{\Phi}^2
\right) \Phi_{ \{ \mu \} \,\{ i\}}=0 \;.\label{emh} \eea The collective
indices $\{ \mu \} \in {\cal R}_{SO(1,4)}$ and $\{ i \} \in {\cal
R}_{SO(5)}$ label irreducible
representations of the ${SO(1,4)}\times SO(5)$ subgroup of the $SO(1,9)$ 
Lorentz group in $D=10$, and run over the 
spectrum of
type IIB string excitations in flat space.  As explained above, 
this guarantees
the right behavior of the KK spectrum in the rigid limit.  The form of
(\ref{emh}) is fixed by Lorentz covariance, while the spectrum of
``masses" $M_\Phi^2$ describing the coupling of a given field $\Phi_{
\{ \mu\}\,\{i\}}$ to the curvature and RR 5-form flux should be ideally
determined by requiring BRS invariance of superstring 
propagation on AdS$_5\times S^5$.

The ten-dimensional field $\Phi_{ \{ \mu \}\,\{i\}}$ can be expanded
in $S^5$-spherical harmonics:
\be
\Phi_{ \{ \mu
\}\,\{i\}}(x,y)=\sum_{[k,p,q]} {\cal X}^{[kpq]}_{\{\mu \}}(x)\,
{\cal Y}^{[kpq]}_{\{i\}}(y)
\;,
\label{harm}
\ee
with $x, y$ coordinates along AdS$_5$ and $S^5$
respectively. The sum runs over a set (to be determined below) of
allowed representations of the
$S^5$ isometry group\footnote{Throughout, we will 
indistinguishly refer to this group as  $SO(6)$ or $SU(4)$ and
exclusively use $SU(4)$ Dynkin labels.}
$SO(6)\approx SU(4)$ characterized by their
$SU(4)$ Dynkin labels
$[k,p,q]$. Finally the 
(generalized) spherical harmonic functions
${\cal Y}^{[kpq]}_{\{i\}}(y)$ are eigenfunctions of the Laplacian:
\bea
\nabla_{S^5}^2 \, {\cal Y}^{[kpq]}_{\{i\}} &=& -\ft{1}{R^2}\left(C_2\left[
SU(4)\right]-C_2\left[ SO(5)\right]\right) \,{\cal Y}^{[kpq]}_{\{i\}}
\;,
\eea
with $C_2[ G]$ standing for the second Casimir of the group $G$.

The aim of this section is to derive the spectrum of harmonics
${\cal Y}^{[kpq]}_{\{i\}}(y)$ entering the expansion (\ref{harm})
using and extending the standard group theory techniques, see
\eg~\cite{deboer,ms} for details. 

\subsection{KK spectrum}
\label{IIBKK}

Let us denote by ${\cal
R}_{SO(5)}$ a representation of $SO(5)$ appearing in the decomposition of 
the spectrum of type IIB string excitations  
on $T^5$ under $SO(1,4)\times SO(5)\subset
SO(1,9)$.  Each $SO(5)$ representation ${\cal
R}_{SO(5)}$ can be associated to a tower of KK descendants on $S^5$ belonging 
to representations of $SO(6)$ that contain ${\cal
R}_{SO(5)}$ in the decomposition $SO(5)\subset SO(6)$.
Denoting by $[m,n]$ the $SO(5)$ Dynkin
labels, the set of $[k,p,q]$ in the KK towers (\ref{harm})
is explicitly given by
\bea
{\rm KK}_{[m,n]} &\equiv& \;\;
\sum_{r=0}^{m} \;\;
\sum_{s=0}^{n} \;\;
\sum_{p=m-r}^\infty \;\;
\left[r\pls s,\,p,\,r\pls n\mis s\right]
\non
&&
+\sum_{r=0}^{m-1} \;\;
\sum_{s=0}^{n-1} \;\;
\sum_{p=m-r-1}^\infty \;\;
\left[r\pls s\pls1 ,\,p,\,r\pls n\mis s\right]
\;.
\label{KK56}
\eea
For instance, the KK towers arising from the lowest $SO(5)$
representations are given by
\bea {\rm KK}_{{\bf 1}} &\equiv& {\rm
KK}_{[0,0]} ~=~ \sum_{p\ge0} \; [0,p,0] \;, \non {\rm KK}_{\bf 4}
&\equiv& {\rm KK}_{[0,1]} ~=~ \sum_{p\ge0} \;\left( [0,p,1]+ 
[1,p,0] \right) \non {\rm KK}_{{\bf 5}} &\equiv& {\rm KK}_{[1,0]} ~=~
 \sum_{p\ge0} \;\left( [0,p\pls1,0]+ [1,p,1]\right) \;,\qquad
\mbox{etc.} \;.  \eea

For our subsequent analysis, it is convenient to observe that, if the 
(in general reducible) $SO(5)$ representation 
${\cal R}_{SO(5)}$ itself may be lifted to a representation
$\widehat{{\cal R}}_{SO(5)}$ of $SO(6)$, its
KK tower takes the very compact form
\bea
{\rm KK}_{{\cal R}_{SO(5)}} &=&
\sum_{n=0}^\infty \; [0,n,0] \times \widehat{{\cal R}}_{SO(5)} \;,
\label{KKR}
\eea
where the r.h.s.\ now describes a tensor product in $SU(4)$. It is
obvious, that the states on the r.h.s.\ contain the states of ${{\cal
R}_{SO(5)}}$ under the breaking of $SO(6)$ to $SO(5)$. With some more
effort and using the explicit form of the KK tower~\Ref{KK56} the
reader may convince himself that these and only these representations
satisfy the embedding condition (\ref{KK56}).

As an illustration, one can explicitly verify this for the first few
$SO(5)$ representations whose KK towers are given above:
\bea
{\rm KK}_{\bf 1} &=&
\sum_{n=0}\; [0,n,0]   \times {\bf 1}  \;,  \nn\\
 {\rm KK}_{{\bf 1}+{\bf 5}} & = &
\sum_{n=0}\; [0,n,0]   \times {\bf 6}  \;,  \nn\\
{\rm KK}_{\bf 4} & = &
\sum_{n=0}\;[0,n,0] \times {\bf 4_s} ~=~
\sum_{n=0}\;[0,n,0] \times {\bf 4_c} \;,
\label{TTT}
\eea
where the l.h.s.\ refer to $SU(4)$ tensor products.\footnote{Recall 
that the
representations ${\bf 4_{s,c}}$, ${\bf 6}$ carry $SU(4)$ Dynkin
labels $[100]$, $[001]$, and $[010]$, respectively.}
The last
example in \Ref{TTT} shows that the presentation of a given KK tower may
not be unique if ${\cal R}_{SO(5)}$ admits different lifts to
$SO(6)$. Nevertheless the full KK tower is unambiguously determined. 
Summarizing the spectrum of harmonics can be found by
first lifting ${\cal R}_{SO(5)}$ to $SO(6)$ and then tensoring it
with $[0n0]$.

Let us now apply this analysis to the massive string spectrum
\Ref{tn}. These states have been organized according to the little
group $SO(9)$. Following the above algorithm, we decompose
this spectrum under $SO(5)\times SO(4)$ and lift $SO(5)$ to $SO(6)$
according to \Ref{KK56}. This uniquely yields the decomposition of the full 
spectrum of KK towers under $SO(6)\times SO(4)$,
corresponding to the product of the
$S^5$ isometry group and the subgroup $SO(4)\subset SO(4,2)$
of AdS$_5$ isometries.

The string spectrum on AdS$_5 \times S^5$ is 
organized in generically long multiplets $\CA^{\Delta}_{[k,p,q](j,\Bj)}$ of
the AdS$_5$ supergroup $SU(2,2|4)$. HWS are characterized by 
their $SU(2)_L\times SU(2)_R$ spins $(j,\Bj)$ and $SU(4)$ Dynkin labels
$[k,p,q]$, that determine the dimension of the supermultiplet to be 
${\rm dim}(\CA^{\Delta}_{[k,p,q](j,\Bj)}) = 2^{16}\times {\rm dim} 
[k,p,q]_{(j,\Bj)}$, and by their scaling dimension
$\Delta = \Delta_{0} +\gamma$ that satisfy unitarity lower bounds \Ref{unitarity} in 
terms of the remaining quantum numbers. The $SU(4)\times SO(4)$
content of long supermultiplets\footnote{Unless otherwise stated, we 
will always indicate the bare dimension $\Delta_{0}$ as a superscript.}  
can be read off from 
\bea \CA^{\Delta_{0}}_{[k,p,q](j,\Bj)} &\equiv&
\hat{T}^{(2)}_1 \times [k,p,q]_{(j,\Bj)}^{\Delta_0 - 2} \;, \la{multA}
\eea
where $\hat{T}^{(2)}_1$ can be identified with 
the by-now-famous (long) Konishi multiplet \cite{andfer,ffz,bkrskon}, 
whose HWS is a scalar singlet of bare scaling dimension 
$\Delta_{0}=2$. Its $2^{16}$ components arise 
from the unconstrained action of 16 supercharges $Q$, \ie
\bea
 \hat{T}^{(2)}_1&=& (1+Q+Q\wedge Q+\ldots)\times [000]_{(0,0)}^{2}
 \;, \quad 
 Q ~=~ [100]^{\frac12}_{(\frac12 ,0)}+[001]^{\frac12}_{(0,\frac12)} \;.
 \label{susys}
\eea %
The quantum numbers of $\hat{T}^{(2)}_1$ were displayed in
tables~1--15 of \cite{andfer}. For convenience of the reader we
collect these states in table~\ref{Kon13}
after the bibliography.\footnote{An extra singlet
$(\theta^4\bar{\theta}^8)_{\ell+6}$ has been included in order to correct
a misprint in table~4 of \cite{andfer}.} Given the conformal dimension
of the HWS the dimensions of its (super)-descendants can be read off from
(\ref{susys}) by assigning $\Delta_{Q}=\ft12$ to the supercharges
$Q$. The conformal dimension $\Delta$ of the HWS should be fixed by
quantizing the superstring fluctuations around the AdS vacuum. In the
limit where the HS symmetry is restored, a large fraction of the
spectrum of dimensions $\Delta(\lambda = 0) =\Delta_0$ is expected
to saturate 
unitary bounds where long multiplets \Ref{multA} become reducible and
split into several semishort or BPS multiplets~\cite{dolosb}. We give
a brief review on this multiplet shortening in
appendix~\ref{app:shortenings} where protected $\frac12$-BPS multiplets
corresponding to the `massless' supergravity states are also
discussed.

Returning to the spectrum of \Ref{tn}, we note that at level
$\ell=1$, the $SO(5)\times SO(4)$ content of $T_1$ exactly
coincides with the Konishi multiplet~\Ref{susys} upon breaking
$SO(6)$ down to $SO(5)$. I.e.\ we are precisely in the situation
discussed in \Ref{KKR}. Hence, we can immediately write down the
entire KK tower at $\ell=1$ organized in supermultiplets
\Ref{multA} as
\bea {\cal H}_1 &=& \sum_{n=0}^\infty \; [0,n,0]^{n}_{(0,0)}
~\times~ \hat{T}^{(2)}_1   ~=~ \sum_{n=0}^\infty \;
\CA^{2+n}_{[0,n,0](0,0)} \;. \label{H1} \eea
Unitarity \Ref{unitarity} requires that $\Delta\ge 2\pls n$.  
The assignment
$\Delta_0=2+n$ seems {\it ad hoc} at this level. We shall justify 
this choice later on.
For string excitations at higher levels, the situation is
more involved. Recall that for $\ell>1$ the flat space
spectrum was given in \Ref{tn} as a product of $T_1$ with the
$SO(9)$ representation $({\rm vac}_\ell)^2$. In order to obtain
the
KK tower in closed form \Ref{KK56}, using the above result for
$T_1$, it remains to lift $({\rm vac}_\ell)^2$ to $SU(4)\times
SO(4)$. At level $\ell=2$, this may be achieved via 
\bea
{\bf 9} &\to&
[0,1,0]^1_{(0,0)}+ [0,0,0]^1_{(\frac12,\frac12)}-[0,0,0]^{2}_{(0,0)}\;.
\label{9lift}
\eea
This requires some comments. Note first that using this formal
lift, the entire KK tower at level $\ell=2$ may be written
as
\bea
{\cal H}_2 &=&
\sum_{n=0}^\infty \; [0,n,0]^{n}_{(0,0)}  ~\times~
\hat{T}^{(2)}_1  ~~\times \left(
 [000]^2_{2(0,0)+(0,1)+(1,0)+(1,1)} +[000]_{(1,1)}^2+ [020]^2_{(0,0)}
\vphantom{_{\frac12}^2}
\right.
\nn \\
&& \left.
 +[101]^2_{(0,0)}
+2\cdot [010]_{(\frac12,\frac12)}^2
-2\cdot [000]_{(\frac12,\frac12)}^{3}
 -2\cdot[010]_{(0,0)}^{3}
 + [000]^{4}_{(00)}
\right)\label{H2}
\;.
\eea
It may now be verified that the formal negative multiplicities in the
last factor of this expression do not lead in fact to any states of
negative multiplicity when multiplied with the infinite sum.  E.g.\
the negative states $[010]$ at $n=0$, $\Delta=3$ are precisely
cancelled by the
corresponding states at $n=1$, etc.  Note that this cancellation
uniquely fixes the relative values of $\Delta$ for the HWS in the
lift \Ref{9lift}. Equation \Ref{H2} thus yields a perfectly
sensible result for the KK tower at level
$\ell=2$.

Higher levels allow similar decompositions
\bea {\cal H}_\ell &=& \sum_{n=0}^\infty \; [0,n,0]^{n}_{(0,0)}
~\times~ \hat{T}^{(2)}_1 ~\times~ ({\rm \widehat{vac}}_\ell)^2 \;,
\label{kk} \eea
where ${\rm \widehat{vac}}_\ell$ denotes the required lift of
the flat space spectrum \Ref{tn} to $SU(4)\times SO(4)$. Although
increasing with the level,
ambiguities in the lift of $SO(5)$ to $SO(6)$ affect only the conformal 
dimensions: the
$SO(6)\times SO(4)$ decomposition of the KK towers is
uniquely fixed following our procedure. 

Of particular interest in our subsequent discussions will be the
sector of string states originating from the completely
symmetric $SO(9)$ representations $[\ell-1,0,0,0]$ appearing in 
${\rm vac}_\ell$.
Adopting the vector lift formula
(\ref{9lift}) for these tensors one finds: \be [\ell-1,0,0,0] \to
\sum_{t = 0}^{\ell-1}
[0,\ell - t  -1,0]_{(\frac{t}{2},\frac{t}{2})}^{\ell}-\ldots
\;,
\label{ells} \ee with dots standing for higher energy states. 
Plugging this in (\ref{kk}) one arrives at a remarkably simple formula
for the scaling dimensions:\footnote{In global coordinates 
corresponding to SYM theory on $R^{+}\times S^{3}$ where radial 
quantization exposes the state - operator correspondence, $\Delta$ 
can be identified with the `energy' i.e. with the eigenvalue of the 
generator of time translation $H= P_{0} + K_{0}$.}    
\be 
\Delta_0 ~=~ 2\ell+n \;,
\label{onshell}
\ee for the corresponding HWS in the superstring spectrum.  In the
rest of the paper we will focus on this particularly interesting
sector of the string spectrum and test the validity of
(\ref{onshell}).  In particular the two extreme cases
$s=t+\bar{t}=2(\ell-1)$ and $t=\bar{t}=0$ will be relevant for our
discussions of HS currents and would-be $\ft14$-BPS states
respectively. We stress that (\ref{onshell}) is expected to hold
only for string states coming from (\ref{ells}), while the remaining
states in $\widehat{\rm vac}_{\ell}$, the first appearing at $\ell=3$,
should be dealt with separately.

 The commensurability of the contribution to
(\ref{onshell}) from the string level $\ell$ with the one from the KK harmonic
$n$ is consistent with our expectation that $R\approx \ap$ at $\lambda
=0$. Semiclassical formulae in inverse powers of $\sqrt{\lambda}$ valid for large $s$ (often
denoted by $S$) and/or $n$ (often denoted by $J$) have been obtained
for string solitons in AdS$_5\times S^5$ by worldsheet methods
\cite{strisol} that may be taken as a hint in extending (\ref{onshell}) beyond the HS symmetric point. Clearly (\ref{onshell}) is expected to only hold for the states appearing in (\ref{ells}). At higher level, 
$\ell\ge 3$, other kinds of states are present. In particular, it is well known that $\Delta$ grows as $\sqrt{\ell}$ for states maximizing it
\cite{mz}.

Finally, let us turn our attention on ${\cal H}_0$ that collects the
KK descendants of massless type IIB supergravity. States belonging to
${\cal H}_0$ organize in $\ft12$-BPS multiplets ${\cal B}{\cal
B}^{\frac12,\frac12}_{[0,n,0]{(0,0)}}$ of $SU(2,2|4)$
(which we explicitly give in table~\ref{Tsugra} below)
\bea {\cal H}_0&=&\sum_{n=2}^{\infty}\; {\cal B}{\cal
B}^{\frac12,\frac12}_{[0,n,0]{(0,0)}} \;. \label{H0} \eea
 The omitted $n=1$ and $n=0$ multiplets correspond to the
non-propagating singleton multiplet and AdS vacuum respectively.
Supergravity KK recurrences have been extensively studied in the past starting
with \cite{gm}. BPS multiplets are  built by acting with half the
supersymmetries on a chiral primary with conformal dimension
$\Delta_0$ saturating the unitary bound $\Delta_0
[0,n,0]_{(0,0)}=n$. According to holography they are
 associated to SYM multiplets starting with the chiral primaries
${\rm Tr}\, \phi^{(i_1}\dots\phi^{i_{n})}$. In particular, the
$n=2$ ground floor in the KK tower accounts for the $2^8$ degrees
of freedom of type II supergravity as required.

\subsection{Exploiting higher spin symmetry on AdS$_5$}
\label{HSgauge}

As we mentioned in the previous section supersymmetry determines the
spectrum of conformal dimensions (and therefore of ``masses"
$M_\Phi^2$'s) inside a supermultiplet once $\Delta$ is known for the
HWS or for any other component.  This reduces the task of computing
$M^2_\Phi$ to fix $\Delta$ for a single component inside each
supermultiplet.  In this section we exploit the higher spin symmetries
of ${\cal N}=4$ SYM at $\lambda =0$ in order to justify the bare
conformal dimensions $\Delta_0$ assigned in the previous section to KK
descendants of higher string excitations on AdS$_5\times S^5$. The new
symmetries manifest themselves in the shortening of long multiplets
saturating unitary bounds and containing particles that become
massless in the free SYM limit.

Following \cite{ffz} we define the AdS$_5$ mass of a given
field belonging to the $SO(4,2)$ representation\footnote{$SO(4,2)$ 
`Dynkin labels' $[j,\Delta,\Bj]$ are related to $SO(6)$ Dynkin labels
$[a,b,c]$ by $\Delta = - b$, $j = {\ft12} (a+c)$, $\Bj = {\ft12} (a-c)$.} 
$D(\Delta,j,\Bj)$
 by
\be m^2=C_2\left[ SO(4,2)\right]-m_0^2
=\Delta(\Delta-4)-\Delta_{\rm min} (\Delta_{\rm min} -4)\;, \ee with
\be C_2\left[
SO(4,2)\right]=\Delta(\Delta-4)+2j(j+1)+2\Bj(\Bj+1) \;,
\label{c2}\ee
 and $m_0^2$ a shift chosen in such a way that massless
 representations on AdS saturate one of the following 
 bounds:\footnote{Obviously $\Delta_{\rm min} = 1$ for scalar
fields different 
from the identity operator.}
\bea
\Delta &\geq&
\Delta_{\rm min}~\equiv~ 2+j+\Bj\quad\quad\;\,
j,\Bj\neq 0\;,\nn\\
 \Delta &\geq& \Delta_{\rm min}~\equiv~ 1+j\quad\quad\quad~~~
j\neq 0,\Bj=0\;,\nn\\
\Delta &\geq& \Delta_{\rm min}~\equiv~ 1+\Bj\quad\quad\quad~~~
j= 0,\Bj\neq0\;,\nn\\
\Delta & \geq & \Delta_{\rm min}~\equiv~ 0
 \quad\quad\quad\quad~~~~~
j=\Bj=0 \;.\label{dmin}
\eea
We list in table~\ref{m2} some relevant cases. In particular the
$SO(6)$ gauge vectors with $\Delta=3$ and the graviton with $\Delta=4$ account for the massless modes of ${\cal N}=8$ supergravity i.e. $\ell=0$.
\begin{table}[bth]
\centering
\begin{tabular}{|l|lll|}
\hline
 type & SO(4) & $\Delta_{\rm min}$ & $m^2$ \\
\hline
 scalar & (0,0) &   0
   & $\Delta(\Delta-4)$ \\
 vector & $(\ft12,\ft12)$ & $3$ & $(\Delta-1)(\Delta-3)$\\
 Ant. tensor & $(1,0)$, $(0,1)$ & $2$ & $(\Delta-2)^2$\\
 metric & $(1,1)$ & $3$ & $\Delta(\Delta-4)$\\
 tensors& $({s\over 2},{s\over 2})$ &$2+s$& $(\Delta-2-s)(\Delta-2+s)$\\
       & $(s,0)$, $(0, s)$ &$1+s$& $(\Delta-1-s)(\Delta-3+s)$\\
\hline
\end{tabular}
\caption{$m^2$ in AdS$_5$. $s\neq 0$ is understood.}
\label{m2}
\end{table}

We are interested in the free SYM limit that exhibits an enhanced
$hs(2,2|4,\mathbb{R})$ higher spin symmetry.  According to holography
this should correspond to higher spin particles in AdS$_5$ becoming
massless as $\lambda \to 0$. Unlike the $SO(6)$ gauge vectors and the
metric tensor of supergravity, the gauge particles realizing this
higher spin symmetry sit in semishort (rather than $\frac12$-BPS)
multiplets and can become massive by `eating' Goldstone particles when
the 
dilaton takes a VEV. The boundary conformal field theory counterpart
of this quantum effect is the emergence of anomalous
dimensions\footnote{Semishort \cite{aeps} and 1/4 BPS multi-trace
operators \cite{dhhhr} dual to multi-particle states at threshold
remain short if there are no `state/operators' to pair with.}  showing
up in logarithmic behaviors at short distances
\cite{bkrslog,bkrsandim}. The higher spin currents are indeed
conserved only in the strict $\lambda=0$ limit where the $SO(4,2)$
unitary bound $\Delta_0=2+j+\Bj$ be saturated.  It is easy to identify
these currents in the string spectrum (\ref{kk}). They correspond to
multiplets built out of the highest spin components $({\rm
vac}_\ell)^2=[000]_{(\ell-1,\ell-1)}+\ldots$, described in
(\ref{ells}).  Masslessness in AdS$_5$ requires that the bound
$\Delta_0=2\ell= 2 +s$ is saturated by the HWS, confirming the {\it ad
hoc} assignment (\ref{onshell}). The corresponding long multiplets
decompose into a sum of BPS/semishort ones according to
(\ref{decomposition})
\be
{\cal A}^{2\ell}_{[000]{(\ell-1,\ell-1)^*}}\approx {\cal C}{\cal
C}^{{1},{1}}_{[000]{(\ell-1,\ell-1)^*}}+\ldots \;.
\ee 
Higher spin conserved currents sit in ${\cal C}{\cal
C}^{{1},{1}}_{[000]{(\ell-1,\ell-1)^*}}$.  Here and below we denote
massless representations by $(j,\Bj)^*$ defined after the subtraction
of the gauge degrees of freedom $(j-\frac12,\Bj-\frac12)$.  The
details can be found in \cite{dolosb} or in
appendix~\ref{app:shortenings}.  The content of gauge multiplets is
displayed in table~\ref{C11} in the appendix.  Taking into account
that all fields in this multiplet satisfy massless field equations,
the total dimension of this multiplet is given by $256\, (2\ell+3)$.
The spectrum agrees with that of the massless higher spin theory
studied in \cite{sezsun5} and realized in terms of a single massless
multiplet of the higher spin algebra $hs(2,2|4)$.  Linearized field
equations for these fields in AdS$_5$ have been worked it out in
\cite{sezsun5}.  The remaining multiplets in the decomposition include
the Goldstone particles needed for the Higgsing of the higher spin
gauge symmetry at $\lambda \neq 0$.

KK descendants with $n>0$ of conserved currents sitting inside ${\cal
A}^{\Delta_0}_{[0n0]{(\ell-1,\ell-1)}}$ do no longer satisfy the
$SO(2,4)$ unitarity bound but they rather realize the
$SU(2,2|4)$ bound (\ref{unitarity}):
\ben
\Delta_0 ~=~ 2\ell+n \;.
\een
This condition generalizes to AdS$_5\times S^5$ the massless
bound in AdS$_5$.  Long multiplets starting with HWS saturating this bound 
decompose again in terms of semishort
multiplets according to (\ref{decomposition}). 
The pattern of shortenings in (\ref{decomposition}) shows 
similarities but at the same time some differences for the Konishi 
multiplet, its HS `relatives' and their KK 
excitations. The celebrated ${\cal N}=4$ Konishi anomaly \cite{konan} 
translates into a violation of a linear type shortening condition 
\cite{bkrskon}. 
The KK excitations of the Konishi multiplet are 'violated' by a generalized  
Konishi anomaly that can be easily deduced from the results of \cite{wittetal} 
by observing that ${\cal N}=4$ SYM is a very special ${\cal N}=1$ SYM theory 
with three chiral multiplets in the adjoint.   
The role of this ${\cal N}=4$ generalized  Konishi anomaly in the mixing of BMN 
operators \cite{bmn} with two-impurities will be analyzed in \cite{prep}. 
Finally, violation of HS currents in ${\cal N}=4$ SYM requires
Goldstone 
multiplets with fermionic HWS that have so far been largely unexplored.

We would also like to stress once more that the full
superstring spectrum 
decomposes into infinitely many (generically `massive') HS multiplets. 
In particular 
the first Regge trajectory consists of a single `massless' HS gauge 
fields dual, in the holographic sense, to twist-two bilinear currents together 
with their superpartners.
The first KK recurrence and (part of) the Goldstones belong to two 
different and massive HS multiplets dual to twist-three trilinear operators. 
At 
the next level one finds, in addition to the second KK recurrence and 
the remaining Goldstones new massive HS multiplets all of 
which are dual to twist-four quadrilinear operators. Proceeding to
higher levels is straightforward in principle but requires a better understanding of massive HS multiplets.

\section{The spectrum of free ${\cal N}=4$ SYM theory}
\label{gauge}

In this section we derive the spectrum of single
trace operators in ${\cal N}=4$ SYM theory with group $SU(N)$ at large $N$. 
Single trace operators
are associated to `words' built with the `alphabet' consisting of the `letters'
\footnote{We adhere to the nomenclature introduced by A.~Polyakov although 
the terms 'necklaces' 
(for 'words') and 'beads' (for 'letters') sound more romantic and to some 
extent ('cyclicity') more appropriate.} \cite{polpol,kutlar}
$\phi^i$, with $i=1, \dots, 6$, $\lambda_{\alpha}^A$ and 
$\lambda_{\dot\alpha A}$, with $A=1, \ldots,
4$ and $\alpha,\dot{\alpha}=1,2$,
$F_{\mu\nu}$, with $\mu=0, \dots, 3$, and derivative thereof. 
As usual 
$V_{\alpha\dot{\alpha}}= V_\mu \sigma^\mu_{\alpha\dot{\alpha}}$,
$A_{\alpha\beta}= \ft12 A_{\mu\nu}\sigma^{\mu\nu}_{\alpha\beta}$ and 
similarly for $A_{\dot{\alpha}\dot{\beta}}$.

The quantum numbers of
a given operator at $\lambda=0$ can be read off from those of the
building letters collected in table~\ref{tab:letters}. 
\begin{table}[hbt]
\centering
\begin{tabular}{l|lll}
Field & SU(4) & $(j,\Bj)$ & $\Delta$   \cr\hline
 $\phi^i$  & [010] & (0,0) & 1   \cr
 $\partial_{\alpha\dot{\alpha}}$ & [000] & $(\ft12,\ft12)$ & 1  \cr
 $\lambda_{\alpha}^A$ & [100] &
$(\ft12,0)$ & $\ft32$   \cr $\lambda_{\dot{\alpha}A}$ & [001] &
$(0,\ft12)$ & $\ft32$   \cr $F_{\alpha\beta}$ & [000] & $(1,0)$ &
2  \cr $F_{\dot{\alpha}\dot{\beta}}$ & [000] & $(0,1)$ & 2
\end{tabular}
\caption{${\cal N}=4$ SYM multiplet}
\label{tab:letters}
\end{table}

\subsection{Enumerating SYM operators: Polya(kov) Theory}

Enumerating SYM states can look at first sight 
a rather tedious exercise but it can be conveniently handled by 
resorting to 
Polya theory \cite{polya}. This idea was applied by A.~Polyakov in   
\cite{polpol} to the counting of gauge invariant operators 
made out of bosonic `letters'. Here we extend his results and 
enumerate, rather than simply count, gauge invariant words including 
fermionic letters. For future reference, we work out the counting 
from a more general 
perspective
than what we really need. This section is written in a self-consistent
form and can be read independently of the rest of the paper. 

Lets start by briefly reviewing the basics of Polya theory. Consider a set
of words $A,B,\ldots $ made out of $n$ letters chosen 
within
the alphabet $\{ a_i \}$ with $i=1,\ldots p$. Let $G$ be a group action 
defining the equivalence 
relation $A\sim B$ for $A=g B$ with $g$ an element of $G\subset S_n$.  
Elements $g\in S_n$ can be divided into 
conjugacy classes $[g]=(1)^{b_1}\ldots (n)^{b_n}$, 
according to the numbers $\{b_k(g)\}$ of cycles of length $k$.
Polya theorem states that the set of inequivalent words are generated by the 
formula:
\bea
P_{G}(\{ a_i \}) &\equiv& 
{1\over |G|}\, \sum_{g\in G}\; \prod_{k=1}^n\; 
(a_1^k + a_2^k\ldots + a_p^k)^{b_k(g)} \;.
\label{pg}
\eea
In particular, for $G=Z_n$, the cyclic permutation subgroup of $S_n$,
the elements $g\in G$ belong to one of the conjugacy classes
$[g]=(d)^{n\over d}$ for each divisor $d$ of $n$.  The number of
elements in a given conjugacy class labeled by $d$ is given by Euler's
totient function $\varphi(d)$, equal to the number of numbers
relatively prime to $n$ (to be pedantic) smaller than $n$. For $n=1$
one defines $\varphi(1) =1 $.  Computing $P_G$ for $G=Z_n$ one finds:
 \be
P_{n}(\{ a_i \})\equiv {1\over n}\, \sum_{d|n} \varphi (d)
(a_1^d + a_2^d+\ldots+ a_p^d)^{n\over d} \;.
\label{pn}
\ee  
The number of inequivalent words can be read off from (\ref{pg}) by simply 
letting 
$a_i\to 1$. 
For instance, the possible choices of `necklaces' with six `beads' 
of two different `colors' $a$ and $b$, are given by
\bea
P_6 (a,b) &=&{1\over 6}\left[(a+b)^6 + (a^2+b^2)^{3} + 2 
(a^3+b^3)^{3} + 2 (a^6+b^6) \right] \nn \\
&=& a^6  + a^5 b  + 3 a^4 b^{2}  + 4 a^{3} b^{3}+ 3 a^{2} b^{4} + a 
b^5 + b^6 \;, \nn
\eea
and the number of different necklaces is $P_6(a=b=1)=14$.

Let now come to our main task: To count the number of gauge invariant
words (single trace operators) in four-dimensional $SU(N)$ YM theories with fields in
the adjoint representation.\footnote{Other gauge groups and/or other
representations for the matter field require a little more effort and
will not be considered here.}  As mentioned at the beginning we will
endeavor to be rather general and consider words made out of the
(on-shell) letters:
\bea
 \partial^s \phi  
\in{\bf n_{s}}^{(s+1)} _{(\frac{s}{2},\frac{s}{2})}\;,&&
\quad 
\partial^s F\in
 {\bf n_{v}}^{(s+2)}_{(\frac{s+2}{2},\frac{s}{2})}
+{\bf n_{v}}^{(s+2)}_{(\frac{s}{2},\frac{s+2}{2})}\;,
\non[1ex]
\partial^s \lambda \in 
{\bf n_{f}}^{s+\frac{3}{2}}_{(\frac{s+1}{2}, \frac{s}{2})}\;,&&
\quad
\partial^s \bar\lambda \in 
{\bf n_{\bar{f}}}^{s+\frac{3}{2}}_{(\frac{s}{2}, \frac{s+1}{2})}
\;,
\label{letters}
\eea
accounting for the elementary YM fields and derivatives thereof.
On-shell conditions have been taken care of by restricting the action of the 
derivatives $\partial^s$ to completely symmetric and 
traceless representation of the $SU(2)_L\times SU(2)_R$ Lorentz group.
Finally ${\bf n_s},{\bf n_f},{\bf n_{\bar{f}}},{\bf n_v}$ refers to the number 
of scalars, spinor left, spinor right
and vector, respectively. In most cases these labels denote (reducible) 
representations of a global symmetry group.

In order to count words one has to take into account the cyclicity
of the trace. 
This is exactly what the cycle index (\ref{pn}) does for us. The only subtleties 
are the arbitrary number of letters in a given word, the infinite number of 
letters in the alphabet and the statistics of the letters. It is easy to convince oneself 
that
\bea
{\cal Z}_{\rm YM}(q) &=& \sum_{n, d|n} {\varphi (d)\over n}\,  
\left[{\cal F}^{(d)}_s(q) + {\cal F}^{(d)}_v(q) 
- {\cal F}^{(d)}_f(q) - {\cal F}^{(d)}_{\bar{f}}(q)\right]^{n\over d}
\nn \\
&=&
- \sum_d {\varphi (d)\over d} \log[1 - {\cal F}^{(d)}_s(q) - {\cal F}^{(d)}_v(q) 
+ {\cal F}^{(d)}_f(q) + {\cal F}^{(d)}_{\bar{f}}(q)] \;,
\label{pnym}
\eea  
does the job for $U(N)$ at large $N$. For $SU(N)$ one simply eliminates words 
with only one letter, \ie~$n\geq 2$. Finally the sign takes 
care of the right spin statistics. In particular, fermion states will appear at 
half-integer powers of $q$ with an overall negative sign. Incidentally we notice 
that is exactly the way one derives Fermi statistics.

Following \cite{polpol,kutlar} we define the partition functions of each species 
\be
{\cal F}^{(d)}_\Phi (q) = \sum_{s,I} 
q^{d(s + \Delta_\Phi)} (\partial^s \Phi_I)^{d} \;,
\ee
where $I$ is a multi-index encompassing spin and 'flavour' 
in (\ref{letters}) while
 the parameter $q$ traces the contribution
of a given letter to the conformal dimension $\Delta$.

Obviously replacing all $\partial^s \Phi_I$ by 1, one is 
simply counting the on-shell 'letters' of a given species.
Explicitly for the letters in (\ref{letters}) and $D=4$ one finds:
 \bea
 {\cal F}^{{(d)}}_s (q)
  &\to& n_s \sum_{s=0}^\infty (s+1)^2 q^{d(s+1)}=n_s \, 
  {q^{d} (1+q^{d})\over (1-q^{q})^3}\;,\nn\\
  {\cal F}^{{(d)}}_{f} (q)
   &\to& n_{f} \sum_{s=0}^\infty (s+1)(s+2) q^{d(s+\frac32)}
=2\,n_{f} \, {q^{3 d\over 2}\over 
 (1-q^{d})^3}\;,\nn\\
 {\cal F}^{{(d)}}_{{\bar{f}}} (q)
   &\to& n_{{\bar{f}}} \sum_{s=0}^\infty (s+2)(s+1) q^{d(s+\frac32)}
=2\,n_{{\bar{f}}} \, {q^{3 d\over 2}\over 
 (1-q^{d})^3}\;,\nn\\
 {\cal F}^{(d)}_v (q)
 &\to&2\, n_v\sum_{s=0}^\infty (s+1)(s+3) q^{d(s+2)}=
 2\, n_v \, {q^{2d} (3-q^{d})\over 
 (1-q^{d})^3}\;,
 \label{onel}
 \eea
with $n_s, n_f, n_{\bar{f}}, n_v$ defined above.

It is now an easy algebraic exercise to produce explicit formulae for the 
partition functions of pure ${\cal N}$-supersymmetric free YM theories, for which 
\bea
{\cal N}=4 && \quad\quad n_s=6,~n_f=n_{\bar{f}}=4,~n_v=1\;,\nn\\
{\cal N}=2 && \quad\quad n_s=2,~n_f=n_{\bar{f}}=2,~n_v=1\;,\nn\\
{\cal N}=1 && \quad\quad n_s=0,~n_f=n_{\bar{f}}=1,~n_v=1\;.
\eea

Plugging into (\ref{pnym}) and expanding in powers of $q$ up to $\Delta=4$ :
\bea
{\cal Z}_{{\cal N}=4}(q) &=&
21\,q^2 - 96\,q^{\frac{5}{2}} + 361\,q^3 - 
  1328\,q^{\frac{7}{2}} + 4601\,q^4 +\ldots \;,\nn\\
{\cal Z}_{{\cal N}=2}(q)&=& 3\,q^2 - 16\,q^{\frac{5}{2}} + 57\,q^3 - 
  184\,q^{\frac{7}{2}} + 551\,q^4+\ldots \;,\nn\\
 {\cal Z}_{{\cal N}=1}(q) &=&6\,q^3 - 20\,q^{\frac{7}{2}} + 65\,q^4  +\ldots
 \label{n124}  
 \eea
In (\ref{n124}) we have subtracted unphysical modes ${\cal Z}_{\rm unphys}= 
(n_f n_{\bar{f}}-1) q^{3} + 4 q^{4} - 2( n_f+n_{\bar{f}}) q^{7\over 2}$
but not total derivatives (conformal descendants).
    
\subsection{Marginal and relevant SYM operators}
\label{margsym}

Specializing to ${\cal N}=4$ SYM, reinstating the `letters' in (\ref{pnym})
 and expanding up to 
$\Delta=4$ but keeping only integer powers of $q$ we 
are left with the bosonic generating function:
\bea
&& {\cal Z}_{\rm bos}(q)~=~ \ft{1}{2} (\phi_i \phi_j+\phi_i^2)\,
q^2\nn\\
&&+\left[\ft{1}{3}(\phi_i\phi_j\phi_k+2\,\phi_i^3)+
\ft{1}{2} (\bar{\lambda}_{\dot{\alpha} A} \bar{\lambda}_{\dot{\beta} B}-
\bar{\lambda}_{\dot{\alpha} A}^2+{\rm h.c.})+
\lambda_\alpha^{A} \bar{\lambda}_{\dot{\alpha} B} + F_{\mu \nu} \phi_i
+\phi_i\partial_\mu \phi_j\right]q^3\nn\\
&&+\left[\ft{1}{4}(\phi_{i_1}\phi_{i_{2}}\phi_{i_{3}}\phi_{i_4}+
\phi_{i_1}^2\phi_{i_2}^2
+2\phi_{i_1}^4)+(\lambda_\alpha^A+\bar{\lambda}_{\dot{\alpha} A})^2 \phi_i+
(\lambda_\alpha^A+\bar{\lambda}_{\dot{\alpha} A})\partial_\mu
(\lambda_\beta^B+\bar{\lambda}_{\dot{\beta} B})      
\right. \nn\\
&&+ \ft{1}{2}(F_{\mu \nu}F_{\sigma \rho}+F_{\mu \nu}^2)+F_{\mu\nu}\phi^i \phi^j
+\phi_i \phi_j \partial_\mu \phi_k+\phi_i \partial_\mu F_{\mu\sigma}+
\partial_\mu \phi_i  F_{\mu\sigma}\nn\\
&& \left. +\ft{1}{2}(\partial_\mu\phi_i \partial_\nu\phi_j
+(\partial_\mu\phi_i)^2))+\phi_i \partial_\mu\partial_\nu \phi_j\right]q^4+\ldots
\label{partition4}
\eea 
The sum over Lorentz and flavor indices is as before implicitly understood. As 
mentioned before (\ref{partition4}) is valid for $SU(N)$ at large $N$ and 
$\lambda=0$. Single-letter words have been discarded.

\begin{table}[th]
\centering
{\footnotesize
\begin{tabular}{|llll|}
\hline
 $\Delta$ & $(j,\Bj)$   & $SU(4)$ & ${\cal O}$ \cr \hline
 2  & (0,0) & [0,0,0]+[0,2,0]={\bf 1}+{\bf 20}
 & ${\rm Tr}\, \phi^{((i_1} \phi^{i_2))}$ \cr
 \hline
 3  & (0,0) & [0,1,0]+[0,3,0]={\bf 6}+{\bf 50}
 & ${\rm Tr}\, \phi^{((i_1}\phi^{i_2} \phi^{i_3))}$  \cr
    & (0,0)
& [0,0,2]+[0,0,2]={\bf 10}$_s$+{\bf 10}$_c$
   & ${\rm Tr}\, \phi^{[i_1}\phi^{i_2} \phi^{i_3]}$ \cr
  & (0,0)
  & [2,0,0]+[0,0,2]={\bf 10}$_s$+{\bf 10}$_c$
   & ${\rm Tr}\, \lambda_\alpha^{(A}\lambda^{B)\alpha}+{\rm h.c.}$
   \cr
& (1,0) & [0,1,0]={\bf 6}
 & ${\rm Tr}\, F_{\alpha\beta}\,\phi^{i}$  \cr
& (1,0)
 & [0,1,0]= {\bf 6}
   & ${\rm Tr}\, \lambda_{(\alpha}^{[A}\lambda^{B]}_{\beta)} $
   \cr
& $(\ft12,\ft12)^*$ & [1,0,1]={\bf 15}
 & ${\rm Tr}\, \phi^{[i_1}\partial_\mu \phi^{i_2]}$  \cr
& $(\ft12,\ft12)^*$
 &[0,0,0]+[1,0,1]={\bf 1}+{\bf 15}
 & ${\rm Tr}\, \lambda_\alpha^{A}\bar{\lambda}_{\dot{\beta} B}$
   \cr
\hline
 4  &(0,0) & [0,0,0]+[0,2,0]+[0,4,0]=
 {\bf 1}+{\bf 20}+{\bf 105}
 & ${\rm Tr}\, \phi^{((i_1}\phi^{i_2}\phi^{i_3} \phi^{i_4))}$ \cr
  & (0,0)& [0,0,0]+[0,2,0]+[2,0,2]=
   {\bf 1}+{\bf 20}+{\bf 84}
  & ${\rm Tr}\, \phi^{[i_1}\phi^{((i_2]}\phi^{[i_3))} \phi^{i_4]}$  \cr
  & (0,0)& [1,0,1]+[0,1,2]+[2,1,0]=
   {\bf 15}+{\bf 45}$_s$+{\bf 45}$_c$
 & ${\rm Tr}\, \phi^{[i_1}\phi^{i_2}\phi^{((i_3]} \phi^{i_4))}$  \cr
&(0,0) & 2([000]+[1,0,1]+[0,2,0]) =2({\bf 1}+{\bf 15}+{\bf 20})&
${\rm Tr}\, \lambda_\alpha^{[A}\lambda^{B]\alpha} \phi^{i}+{\rm
h.c.} $ \cr & (0,0) & [1,0,1]+[0,1,2]+[2,1,0]=2$\cdot${\bf
15}+{\bf 45}$_s$+{\bf 45}$_c$ & ${\rm Tr}\,
\lambda_\alpha^{(A}\lambda^{B)\alpha} \phi^{i}+{\rm h.c.} $ \cr &
(0,0)& 2[0,0,0]=
   2$\cdot${\bf 1}
 & ${\rm Tr}\, F^2,{\rm Tr}\, F\tilde{F}$
  \cr
&(1,0) & [000]+[1,0,1]+[0,2,0]={\bf 1}+{\bf 15}+{\bf 20} & ${\rm
Tr}\, \lambda_{(\alpha}^{[A}\lambda^{B]}_{\beta)} \phi^{i}$ \cr
&(1,0)  &  [1,0,1]+[2,1,0] =
 {\bf 15}+{\bf 45}$_s$ & ${\rm Tr}\,
\lambda_{(\alpha}^{(A}\lambda^{B)}_{\beta)} \phi^{i} $ \cr & (1,0)
& [0,0,0]+[1,0,1]+[0,2,0]=
  {\bf 1}+{\bf 15}+{\bf 20}
 & ${\rm Tr}\, F_{\alpha\beta}\,\phi^{i_1}\phi^{i_2} $\cr
& $(\ft12,\ft12)$ &2[1,1,1]+2[0,1,0]=
  2$\cdot${\bf 6}+2$\cdot${\bf 64}
 & ${\rm Tr}\,\partial_\mu \phi^{((i_1}\phi^{[i_2))} \phi^{i_3]}$  \cr
&$(\ft12,\ft12)$
 & 4[010]+2[0,0,2]+2[2,0,0]+2[1,1,1]
   & ${\rm Tr}\, \lambda_\alpha^{A}\bar{\lambda}_{\dot{\beta}B} \phi^{i},
 {\rm Tr}\, \lambda_\alpha^{A}\phi^i \bar{\lambda}_{\dot{\beta} B} $ \cr
&&  ~~~~~~~~~=4$\cdot${\bf 6}+2$\cdot${\bf 10}$_s$+2$\cdot${\bf
10}$_c$+ 2$\cdot${\bf 64} & \cr & $(\ft32,\ft12)^* $
 & [2,0,0] ={\bf 10}$_s$
   & ${\rm Tr}\,\lambda_{(\alpha}^{(A}\partial_{\dot{\alpha}\beta} 
   \lambda^{B)}_{\gamma)}$
   \cr
& $(\ft32,\ft12)^* $
  & [0,1,0] ={\bf 6}
 & ${\rm Tr}\, \partial_{\dot{\alpha}(\gamma} F_{\alpha\beta)} \phi^{i}$
   \cr
& (2,0)  & [0,0,0]=
   {\bf 1}
 & ${\rm Tr}\, F_{(\alpha \beta} F_{\gamma\delta)} $\cr
 & $(1,1)^*$
 & [0,0,0]=
   {\bf 1}
 & ${\rm Tr}\, F_{\alpha\beta} F_{\dot{\alpha}\dot{\beta}}$
  \cr
& $(1,1)^*$ & [0,0,0]+[0,2,0]={\bf 1}+{\bf 20}
 & ${\rm Tr}\,  \partial_{(\mu} \phi^{((i_1}\,\partial_{\nu)}\,\phi^{i_2))}$  
 \cr
& $(1,1)^*$ & [0,0,0]+[1,0,1]={\bf 1}+{\bf 15}
 & ${\rm Tr}\,  \lambda_{(\alpha}^A \partial_{\beta)(\dot{\beta}}
  \bar{\lambda}_{\dot{\alpha}) B}  $  \cr
 \hline
\end{tabular}
}
\caption{\small {\cal N}=4 SYM at $\lambda=0$. Brackets denote 
antisymmetrization. Parentheses denote complete
symmetrization when traces cannot appear. Double parentheses denote complete 
symmetrization not excluding traces.}
\label{l0}
\end{table}

Next we organize states in irreducible representations of
$SO(4)\times SU(4)$. The $SU(4)$ content of (\ref{partition4}) can be read
off by letting
\bea
&& \ft{1}{2} (\phi_i \phi_j+\phi_i^2)\to \phi^{((i} \phi^{j))}\nn\\
&& \ft{1}{3}(\phi_i\phi_j\phi_k+2\,\phi_i^3)\to \phi^{((i_1}\phi^{i_2} 
\phi^{i_3))}+
\phi^{[i_1}\phi^{i_2} \phi^{i_3]}\nn\\
&&\ft{1}{2} (\lambda_{\alpha}^A \lambda_{\beta}^B-\lambda_{\alpha}^{A\,2})\to
\lambda_{[\alpha}^{(A} \lambda_{\beta]}^{B)}+\lambda_{(\alpha}^{[A} 
\lambda_{\beta)}^{B]}
\label{map} \;\qquad , \qquad  {\rm ~and~so~on.}
\eea
as always traces are understood.
For instance, the multiplicities of 
states of the form $\phi^1 \phi^1 \phi^1$, 
$\phi^1 \phi^1 \phi^2$, $\phi^1 \phi^2 \phi^3$ in the left hand side of 
(\ref{map}) are $1,1,2$ respectively determining the right hand side. 
This can be confirmed by counting the number of cyclic words of the
corresponding type. In a 
similar
way one can work out the other cases. The final output for   
single-trace operators with $\Delta_0\leq 4$ is given in table
\ref{l0}. 
In building table \ref{l0} we omit total derivatives. 
In addition we discard terms containing
$\partial^2~\phi^i, 
\spartial~\lambda^A,
\spartial~\bar{\lambda}_A, 
\partial^\mu~F_{\mu\nu}$, or $\partial^\mu~\tilde{F}_{\mu\nu},$ 
which vanish along the 'free' field
equations or after imposing Bianchi identities.
In particular, the equations of motion imply the conservation of
all $s=1$ currents at $\Delta=3$ and $s=2$ tensors at
$\Delta=4$.  Special cases of this are
 the $SO(6)$ current at $\Delta=3$ and the stress energy
tensor at $\Delta=4$ realized in supergravity which are the only 
currents whose conservation survives quantum corrections.

\section{String vs.\ gauge theory}
\label{compa}

\subsection{Operators/states with $\Delta\leq 4$ }

Let us now identify the string excitations which are dual to the 
relevant and marginal operators found in the previous section. 
According to (\ref{onshell})
string states with $\Delta\leq 4$ appear only in ${\cal H}_\ell$
with $\ell=0,1,2$. Collecting (\ref{H1}), (\ref{H2}), (\ref{H0}) we
find: 
\bea {\cal H}_{\Delta\leq 4}&\in& \sum_{n=2}^4 {\cal B}{\cal
B}^{\frac12,\frac12}_{[0n0]{(00)}}+ \sum_{n=0}^2 
{\cal A}^{2+n}_{[0n0]{(00)}}+2{\cal A}^{4}_{[000]{(00)}}+
{\cal A}^{4}_{[101]{(00)}}+{\cal A}^{4}_{[020]{(00)}}\nn\\&&
+{\cal A}^{4}_{[000]{(10)}}+{\cal A}^{4}_{[000]{(01)}}+
2{\cal A}^{4}_{[010]{(\frac12,\frac12)}}
+{\cal A}^{4}_{[000]{(11)}} \;.
\label{stringKK}
\eea
The first two terms represent the contributions of supergravity and Konishi 
together with their lowest KK recurrences. 
The remaining multiplets appears at $\ell=2$. 
The genuinely  
long supermultiplets ${\cal A}^{4}_{[000]{(00)}}$ have been studied in 
\cite{bersmix,appss,ans,bks}. 
Beside these, the remaining long multiplets in (\ref{stringKK}) saturate at $\lambda=0$ the unitarity bounds (\ref{unitarity}) and 
split into a sum of BPS/semishort multiplets according to (\ref{decomposition}).
 The resulting spectrum of $\Delta\le4$ string states is collected in tables
\ref{listrep},\ref{listrep0}.
Comparing the multiplicities and quantum numbers in the string/sugra
and SYM tables we find precise agreement! Establishing this one-to-one
correspondence between the spectrum of relevant and marginal
deformations of string and SYM theories provides strong support to the
dimension formula (\ref{onshell}). This remarkably neat formula
remains on a conjectural basis and will require extensions in order to
deal with fermionic and other HWS, not of the type (\ref{ells}),
present at higher levels. However we find it a very appealing starting
point for future analyses of the string spectrum on AdS$_5\times S^5$
especially at the point of minimal radius where the theory drastically
simplify \cite{witjhs,sundb,mikha}.
\begin{table}[h]
\centering
{\footnotesize
\begin{tabular}{|l|l|l|l|}
\hline Multiplet & $\Delta$ & SU(4) & $(j,\bar{j})$ \cr
 \hline & $2$&$[000]$& $(0,0)$\cr ${\cal C}{\cal
C}^{{1},{1}}_{\left[000\right](0,0)^*}$  & $3$ & $[010]$ &
$(0,1)+(1,0)$   \cr &  & $[000]+[101]$ & $(\ft12,\ft12)^*$ \cr
&$4$ & $[000]$ & $(2,0)+(0,2)$\cr && $[000]+[101]+[020]$ &
$(1,1)^*$\cr &&$[002]+[010]$ & $(\ft32,\ft12)^*$ \cr
&&$[200]+[010]$ & $(\ft12,\ft32)^*$ \cr
 \hline
& $3$&$[200]$& $(0,0)$\cr ${\cal B}{\cal
C}^{{1\over4},{3\over4}}_{\left[200\right](0,0)}$& $4$&
$[000]+[101]+[020]$ & $(0,0)$ \cr & & $[210]$ & $(0,1)$\cr &&
$[101]$ & $(1,0)+(0,1)$\cr & & $[111]+[200]+[010]$&
$(\ft12,\ft12)$ \cr
 \hline
&
 $3$&$[002]$& $(0,0)$\cr
${\cal C}{\cal B}^{{3\over4},{1\over4}}_{\left[002\right](0,0)}$
& $4$& $[000]+[101]+[020]$ & $(0,0)$ \cr & & $[012]$ & $(1,0)$\cr
&& $[101]$ &$(1,0)+(0,1)$\cr & & $[111]+[002]+[010]$&
$(\ft12,\ft12)$ \cr
 \hline
${\cal B}{\cal B}^{{1\over4},{1\over4}}_{\left[202\right](0,0)}$ &
$4$ & $[202]$ & $(0,0)$ \cr \hline & $3$ & $[010]$ & $(0,0)$ \cr
${\cal C}{\cal C}^{{1\over2},{1\over2}}_{\left[010\right](0,0)}$ &
$4$ & $2\!\cdot\![101]$ & $(0,0)$ \cr &&
$[111]+2\!\cdot\![010]+[200]+[002]$ & $(\ft12,\ft12)$ \cr &&
$[000]+[101]+[020]$ & $(1,0)+(0,1)$ \cr \hline ${\cal B}{\cal
C}^{{1\over4},{1\over2}}_{\left[210\right]({0},{0})}$ & $4$ &
$[210]$ & $(0,0)$ \cr \hline ${\cal C}{\cal
B}^{{1\over2},{1\over4}}_{\left[012\right]({0},{0})}$ & $4$ &
$[012]$ & $(0,0)$ \cr \hline ${\cal C}{\cal
C}^{{1\over2},{1\over2}}_{\left[020\right](0,0)}$ & $4$ &  $[020]$
& $(0,0)$ \cr \hline \hline ${\cal C}{\cal
C}^{1,1}_{\left[000\right]({1},{1})^*}$ & $4$ & $[000]$ &
$(1,1)^*$ \cr \hline ${\cal C}{\cal
C}^{{1\over2},{1\over2}}_{\left[020\right](0,0)}$ & $4$ &  $[020]$
& $(0,0)$ \cr \hline ${\cal C}{\cal
C}^{{1\over4},{1\over4}}_{\left[101\right](0,0)}$ & $4$ &  $[101]$
& $(0,0)$ \cr \hline${\cal C}{\cal
C}^{{1\over2},{1\over2}}_{\left[010\right](\frac12,\frac12)}$ &
$4$ &
 $[010]$ & $(\frac12,\frac12)$ \cr \hline ${\cal A}{\cal C
}^{1}_{\left[000\right](0,1)}$
& $4$ &  $[000]$ & $(0,1)$ \cr \hline ${\cal C}\!{\cal
A}^{1}_{\left[000\right](1,0)}$ & $4$ &  $[000]$ & $(1,0)$
\cr \hline $2\cdot {\cal A}^4_{\left[000\right](0,0)}$ & $4$ &
$[000]$ & $(0,0)$ \\ \hline
\end{tabular}
 \caption{\small Bosonic modes with $\Delta\le4$ in the
string spectrum at $\lambda=0$.
}
\label{listrep} }
\end{table}

\begin{table}[h]
\centering
{\footnotesize
\begin{tabular}{|l|l|l|l|}
\hline & $\Delta$ & $SU(4)$ & $(j,\Bj)$ \cr   \hline & $2$&$(020]$&
$(0,0)$\cr   ${\cal B}{\cal B}^{{1\over 2},{1\over
2}}_{\left[020\right](0,0)^*}$ &
 $3$& $[002]+[200]$ & $(0,0)$\cr
& & $[010]$ & $(0,1)+(1,0)$   \cr   && $[101]$& $(\ft12,\ft12)^*$
\cr   &$4$ & $2\!\cdot\![000]$ & $(0,0)$\cr   && $[000]$&
$(1,1)^*$ \cr   \hline & $3$&$[030]$& $(0,0)$\cr   ${\cal B}{\cal
B}^{{1\over 2},{1\over 2}}_{\left[030\right](0,0)}$& $4$&
$[012]+[210]$ & $(0,0)$\cr   && $[020]$ & $(0,1)+(1,0)$   \cr   &&
$[111]$& $(\ft12,\ft12)$ \cr   \hline ${\cal B}{\cal B}^{{1\over
2},{1\over 2}}_{\left[040\right](0,0)}$ & $4$&$[040]$& $(0,0)$ \cr
\hline
\end{tabular}
 \caption{\small Bosonic modes with $\Delta\le4$ in the supergravity
spectrum.
}
\label{listrep0} }
\end{table}

\subsection{BPS and semishort multiplets}
\label{bpssemi}

We would like to focus now on string/SYM states that belong to 
``nearly protected''
long multiplets of the AdS supergroup, \ie~multiplets starting
with
a highest weight state $[k,p,q]_{(j,\Bj)}$ saturating unitarity bounds
 at $\lambda=0$.  For simplicity we
restrict ourselves to the series $k-q=j-\Bj=0$. A self-consistent
description of the decomposition of such multiplets into BPS/semishort
multiplets of $SU(2,2|4)$ is presented in
appendix~\ref{app:shortenings} along the lines of \cite{dolosb}.
Here we would like to count multiplets falling in this particularly
interesting class.

We start by looking at the SYM side. From table \ref{tab:letters}
we see
that all letters, and therefore all words at $\lambda=0$, satisfy
the bound
\be
\Delta\geq j+\Bj+k+p+q \;.
\ee
BPS multiplets start with operators saturating this bound with
$j=\Bj=0$.  For operators with $j=\Bj=0$, the effective bound
$\Delta \geq p+2q$ is only saturated by the letters $\phi^i$ and
therefore highest weight states of BPS multiplets are built only out
of $\phi$'s. There are two kinds of BPS multiplets in the SYM
spectrum\footnote{With a slight abuse of terminology, some
semishort multiplets 
are sometimes referred to as 1/8 BPS.}:  $\ft12$-BPS multiplets
associated to the 
familiar CPO family
${\cal B}_{[0,n,0](0,0)}^{\frac12,\frac12}$ and $\ft14$-BPS multiplets
${\cal B}_{[q+2,p,q+2](0,0)}^{\frac14,\frac14}$ appearing in the
decomposition of long multiplets ${\cal A}^{2+p+2q}_{[qpq](0,0)}$ (see
Appendix). This implies that we can count ``nearly protected" long
multiplets starting with a scalar HWS by simply counting
would-be $\ft14$-BPS multiplets \cite{bks}. The $\ft12$-BPS multiplets are dual
to KK descendants of type IIB supergravity on AdS$_5\times S^5$ and
appear with degeneracy one at each $n$. A systematic study of
$\ft14$-BPS states is given in \cite{dhhhr} and a complete
list of multiplicities for $\ft14$-BPS states with $\Delta\leq 12$ can
be found in table 2 of \cite{bks}.

Let us now consider the string dual picture. To start with we consider
long multiplets with $k=q=0$ i.e. ${\cal
A}^{\Delta_0}_{[0,\Delta_0-2,0]{(0,0)}}$. These are the so called ``BMN
multiplets'' extensively studied in \cite{beis}.  Despite their
name, the definition of ``BMN multiplets'' does not necessary require
any BMN limit and they are in representations of the AdS supergroup
rather than that of the pp-wave geometry \cite{ppwaves}.  
The name is motivated by
the fact, proven in \cite{beis}, that in the large $J$ limit
these multiplets contain all BMN operators with two impurities.  In
the dual SYM theory, they correspond to multiplets starting with
operators of the type ${\rm Tr }\, \phi^k\, \phi^{(i_1}\ldots \phi_k\,
\phi^{i_{\Delta_0-2})}$ \cite{beis}.  There are
$\left[{\Delta_0\over 2} \right]$ operators of this type corresponding
to all possible distances between the two $\phi_k$'s (see also
\cite{dhhhr}).  This matches the multiplicity of highest
weight states in this representation in the string side
(\ref{kk}). This can be seen by noticing that there is a single
highest weight state in (\ref{kk}) with quantum numbers $[0,2\ell+
n-2,0]_{(0,0)}^{2\ell+n}$ for any pair $(\ell,n)$. The
multiplicity is then given by all possible choices of $\ell$ and $n$
for a fixed $\Delta=2\ell+n$, which is precisely $\left[{\Delta\over
2} \right]$. 

Another interesting class of nearly protected long multiplets in the
string spectrum is the one that includes the highest spin components
${\cal A}^{2\ell}_{[000](\ell-1,\ell-1)}$ in (\ref{kk}).  As we have
seen in the previous section, states in these multiplets represent the
higher spin conserved currents of $hs(2,2|4)$ and their KK
descendants. Once again it is easy in principle although quite
laborious in practice to construct the corresponding SYM duals ${\rm
Tr} \phi^i \, \partial_{\mu_1}\ldots \partial_{\mu_{2\ell-2}}\, \phi_i
+ \dots$, where $\dots$ stands for fermion and gauge bilinear needed
for 'superprimarity'. There is a single states of this kind for each
even spin $s=2\ell-2$ (the other being superdescendants !) in
agreement with the string result and therefore the same higher spin
algebra is realized in the two sides of the holographic map.

More general $\ft14$-BPS multiplets ${\cal B}{\cal
B}^{\frac14,\frac14}_{[q+2, p, q+2]{(0,0)}}$ appear 
for example in the
decomposition of long multiplets ${\cal A}^{2+p+2q}_{[qpq]{(0,0)}}$
in the string spectrum:
\bea
{\cal H}_\ell&=&
\sum_{n=0} \; [0n0]_{(0,0)}\times
([0,\ell-1,0]_{(0,0)})^2\times \hat{T}^{(2)}_1+\ldots \label{string14}
\eea
Unlike the case $q=0$ (``two impurities" case) discussed above,
(\ref{string14}) cannot be the only source for $\ft14$-BPS states as
can be seen by comparing the output of (\ref{string14}) to the SYM
$\ft14$-BPS multiplicities listed in table 2 of~\cite{bks}. The
mismatches start at $\Delta=8$, indicating that states other than of
the type (\ref{ells}) show up at this level,
cf.~appendix~\ref{app:vac}. Reversing the argument, one can try to use
the SYM results to get some hints on how our mass formula should be
modified in order to account for such states. On the SYM side, the enumeration
of $\ft14$-BPS states is greatly simplified by the aid of Polya
theory. We plan to come back to this issue in the near future.

\subsection{Superprimaries and Eratostene's super-sieve}
\label{suprim}

Clearly the comparison of the spectrum of SYM operators and string
states can be simplified if we organize SYM operators in
supermultiplets. ${\cal
N}=4$ supersymmetry transformations of (constant) parameters 
$\eta^{A}_\alpha,\bar{\eta}^{\dot \alpha}_A$ read
\bea
 &&\delta_Q(\eta) \phi^{i} = 
\bar{\tau}_{AB}^i \eta^{A}_\alpha  \lambda^{B\alpha} 
\;,\nn\\
 &&\delta_Q(\eta)\lambda^{A}_\alpha = 
{1\over 2} F_{\mu\nu} \sigma^{\mu\nu}{}_\alpha{}^\beta \eta^{A}_{\beta} + 
{1\over 2} g_{_{\rm YM}}[\phi^{i},\phi^{j}]\tau_{ij}{}^A{}_B \eta^{B}_{\alpha}
\;,\nn\\
 &&\delta_Q(\eta)\bar\lambda_{A\dot{\alpha}} = 
\bar{\tau}_{AB}^i \eta^{B \alpha}
 \sigma^\mu_{\alpha\dot{\alpha}} D_{\mu} \phi_i 
\;,\nn\\
 &&\delta_Q(\eta) F^{\mu\nu} = i \eta^{A\alpha}
 \sigma^{[\mu}_{\alpha\dot{\alpha}} D^{\nu]}
\bar{\lambda}_{A}^{\dot{\alpha}}
\;,\label{SYMsusy} 
\eea
where $\tau^{AB}_i, \bar{\tau}_{AB}^i$ are $4\times 4$ Weyl blocks of 
Dirac matrices 
in $D=6$, and $D_\mu$ denote covariant derivatives. 

In addition, ${\cal N} =4$ SYM theory is invariant under 
superconformal transformations that simply read 
\bea
\delta_S(\xi) \phi^{i} &=& \bar\delta_Q(\bar\zeta) \phi^{i} 
\;,\nn\\
\delta_S(\xi) \lambda^{A}_\alpha &=&
\bar\delta_Q(\bar\zeta)\lambda^{A}_\alpha + \phi^{i} \tau^{AB}_i\xi_{B\alpha} 
\;,\nn\\
\delta_S(\xi)\bar\lambda_{A\dot{\alpha}} &=&
\bar\delta_Q(\bar\zeta) 
\bar\lambda_{A\dot{\alpha}}
\;,\nn\\
\delta_S(\xi) F_{\mu\nu} &=& \bar\delta_Q(\bar\zeta)F_{\mu\nu} + 
 \xi_A \sigma_{\mu\nu}\lambda^{A} 
\;.\label{SYMsuconf} 
\eea
As indicated the `orbital' part is a susy transformation 
(\ref{SYMsusy}) with $x$-dependent parameters
$\zeta^{A}_\alpha = x_{\alpha\dot{\alpha}}\bar\xi^{A\dot{\alpha}}$, 
$\bar{\zeta}^{\dot \alpha}_A = x^{\dot{\alpha}\alpha}\xi_{A\alpha}$.

Superconformal primaries, \ie~HWS of $SU(2,2|4)$ supermultiplets, are
defined by the condition 
\be
\hat\delta_S {\cal O} \equiv [\xi_A S^A + \bar\xi^A \bar{S}_A, {\cal O}]= 0
\;.\label{suprimo}
\ee
at the origin $x=0$, where the orbital part 
$\delta_{Q}(\zeta) = \bar\delta_{Q}(\bar\zeta) =0$ 
can be neglected. 

{}From (\ref{SYMsuconf}) one may naively conclude that operators made
with only scalar fields be superprimaries. Indeed, this is true at
$\lambda=0$ but the presence of the scalar commutator term in the
supersymmetry transformation of the gaugino suggests that only
completely symmetric (not necessarily traceless) combinations of
scalar fields should remain HWS of (long) SYM multiplets at
$\lambda\neq 0$.  When interactions are turned on the resolution of
the resulting operator mixing and the identification of the
renormalized HWS requires in general a rather laborious analysis
\cite{bersmix,appss}. In some cases, the problem of diagonalizing the
dilation operator \cite{bkps} can be somewhat simplified or at least
systematized by mapping it into the equivalent problem of
diagonalizing the Hamiltonian of an integrable $SO(6)$ or even $SU(2)$
spin-chain \cite{mz,bks}. A supersymmetric version of the spin-chains
of \cite{mz,bks} should be needed in order to describe fermion and
vector impurities.

In principle, at $\lambda=0$, the condition (\ref{suprimo}) completely 
characterizes the HWS and is straightforward to impose using 
(\ref{SYMsuconf}) and their generalization taking care of the presence of 
derivatives 
\bea
\delta_{0S}(\xi) \partial_{\mu_1}\dots\partial_{\mu_k}\phi_i &=& 
{\tau}_i^{AB} \xi_A 
\sigma_{(\mu_1}\partial_{\mu_2}\dots\partial_{\mu_k)}\bar\lambda_B 
\;,\nn\\
\delta_{0S} \partial_{\mu_1}\dots\partial_{\mu_k} \lambda^A &=& 
{\tau}_i^{AB} \xi_B \left(\partial_{(\mu_1} - \ft14 
\sigma_{(\mu_1}\spartial\right) \dots\partial_{\mu_k)}\phi^i 
\;,\nn\\
\bar{\delta}_{0S} \partial_{\mu_1}\dots\partial_{\mu_k} 
 \bar\lambda_A &=& \xi_A \partial_{(\mu_1}\dots\partial_{\mu_{k-1}} 
 F_{\mu_k)\nu}\sigma^\nu 
\;,\nn\\
\delta_{0S}\partial_{\mu_1}\dots\partial_{\mu_k}F_{\mu\nu}  &=& \xi_A 
 \partial_{(\mu_1}\dots\partial_{\mu_k}\sigma_{\mu)\nu} \lambda^{A} 
\;.\label{SYMsuconfder} 
\eea
In practice the situation even for 'words' with few 'letters' soon tends to be 
very intricate. Decomposing the action of the superconformal charges $S^A$, 
$\bar{S}_A$ on a given 
combination of operators (with the same quantum numbers) into irreducible 
representations of $SO(4)\times SU(4)\subset SU(2,2|4)$ turns out to be very 
helpful in 
practical computations. Proceeding this way we have been able to identify 
the HWS states of the supermultiplets containing operators with $\Delta_0\leq 4$ 
{\it viz.}
 \bea 
{\cal Z}_{\Delta_0\leq 4}&\in& 
\sum_{n=2}^{4}{\cal B}{\cal B}^{\frac12,\frac12}_{[0n0](00)}+ 
{\cal C}{\cal C}^{{1},{1}}_{\left[000\right](00)^*}
+
{\cal B}{\cal C}^{{1\over4},{3\over4}}_{\left[200\right](00)}+
{\cal C}{\cal B}^{{3\over4},{1\over4}}_{\left[002\right](00)}
+{\cal B}{\cal B}^{{1\over4},{1\over4}}_{\left[202\right](00)}\nn\\
&& {\cal C}{\cal C}^{{1\over2},{1\over2}}_{\left[010\right](00)}  
+{\cal B}{\cal C}^{{1\over4},{1\over2}}_{\left[210\right]({0}{0})}
+{\cal C}{\cal B}^{{1\over2},{1\over4}}_{\left[012\right]({0}{0})} 
+{\cal C}{\cal C}^{{1\over2},{1\over2}}_{\left[020\right](00)}
 +{\cal C}{\cal C}^{1,1}_{\left[000\right]({1}{1})^*}+
{\cal C}{\cal C}^{{1\over2},{1\over2}}_{\left[020\right](00)}\nn\\&&
+{\cal C}{\cal C}^{{1\over4},{1\over4}}_{\left[101\right](00)}+ 
2\,{\cal C}{\cal C}^{{1\over2},{1\over2}}_{\left[010\right](\frac12\frac12)} 
+{\cal A}{\cal C}^{1}_{\left[000\right](01)} 
+{\cal C}\!{\cal A}^{1}_{\left[000\right](10)} 
+2\,{\cal A}^4_{\left[000\right](00)} \;.
\eea
The nature of the multiplets to which the primaries belongs to is unambiguosly
determined by their scaling bare dimension $\Delta_0$.
Alternatively one can achieve the same goal by filtering the spectrum
by a sort of `Eratostene's super-sieve'. In other words, starting from
the lowest states $[020]_{(00)}^{2}$ and $[000]_{(00)}^{2}$, one can
subtract from ${\cal Z}_{\rm SYM}$ (\ref{pnym}) the contributions of
the full supermultiplets built on top of them and proceed to HWS of
increasing dimension.  Proceeding this way one can identify -- one
after the other -- all HWS as lowest dimension states which are not
(super)descendants of other HWS and decompose ${\cal Z}_{\rm SYM}$ in
supermultiplets of $SU(2,2|4)$. As a bonus of this `Eratostene's
super-sieve' procedure one should be able to recognize the pairing
with Goldstone multiplets of multiplets saturating the unitary bound
in free theory. The absence of the required Goldstone multiplets would
be an uncontroversial indication that a multiplet remain semishort or
1/4 BPS at large $N$. Mixing with multi-trace operators may complicate
the situation at finite $N$.

\section{Conclusions}
\label{conclusions}

Using `standard' KK techniques and making some plausible assumptions
we have derived the spectrum of KK descendants of fundamental string
excitations on AdS$_5\times S^5$. The results can be written in the
deceivingly simple and suggestive form \bea {\cal H}_\ell&=&
\sum_{n,\ell} \; [0n0]_{(0,0)}\times \hat{\cal H}^{\rm
flat}_\ell\;,\label{final} \eea with $\hat{\cal H}^{\rm flat}_\ell$
the on-shell flat spacetime string spectrum at level $\ell$, properly
rearranged in representations of the AdS supergroup $SU(2,2|4)$.
Already the fact that on-shell string spectrum in flat spacetime can
be organized in multiplets of $SU(2,2|4)$ is remarkable.  In
particular a vector of the $SO(9)$ massive little Lorentz group is
naturally lifted to an $SO(1,9)$ vector minus a scalar unphysical
component~\Ref{9lift}. The latter accommodates the $SO(1,3)\times
SO(6)$ quantum numbers displayed in (\ref{final}).  Remarkably, states
with negative multiplicities have been shown to disappear once the
full KK tower of descendants is taken into account.

The result (\ref{final})
generalizes the familiar tower of $\ft12$-BPS supergravity
multiplets at $\ell=0$, decorating it with infinite towers of long
multiplets $\ell\geq 1$ with increasing spins and masses.
Whatever string theory on AdS$_5\times S^5$ is, it should contain
the spectrum of string theory in flat spacetime at the bottom of
its KK tower and the right supergravity content. The spectrum
(\ref{final}) satisfies both of these requirements.

Naive extrapolation to small radius combined with the tight
constraints imposed by the restoration of HS symmetry led us to
conjecture a very simple yet rather effective mass formula that
encompasses a large fraction of string excitations.  More precisely,
bare conformal dimensions in the bulk string theory are assigned in
such a way that the $\lambda\to 0$ limit in the boundary theory
corresponds to the point where masses of most string states saturate
the respective unitary bounds. Based on the resulting mass formula, we
confirm that in this limit the spectrum contains precisely one
massless HS multiplet in the first Regge trajectory \cite{sezsun5}
together with its KK recurrences and Goldstone multiplets in the
second and higher Regge trajectories.  Genuinely massive HS multiplets
also appear.

Using Polya enumerative theory and generalizing the results of A.
Polyakov~\cite{polpol} we have computed the partition function
of gauge invariant single-trace composite operators in free ${\cal
N}=4$ SYM theory.\footnote{Mixing with multi-trace operators is
suppressed in the large N limit relevant to our analysis.} The SYM
spectrum at $\lambda=0$ is tested in some detail against the string
results. In particular, the SYM spectrum of multiplicities 
and quantum numbers of higher spin 
currents and that of marginal and relevant operators are shown to
precisely match with the one of string theory based on our simple 
mass formula. This remarkable agreement may however be slightly deceiving
since we expect the need for an extension of the mass formula, even 
in the limit under consideration, in order to accomodate states which 
are far from any unitary bound \cite{mz}.
Finally, we have discussed how to exploit
superconformal symmetry in order to select HWS at $\lambda=0$.
Alternatively, an easy-to-apply procedure to achieve the same goal has 
been described that resembles Eratostene's sieve for prime numbers.

The string/SYM spectrum displays rich patterns of shortenings of the
type extensively studied in \cite{afsz, dolosb}. The bulk counterpart
of this is a ``grande bouffe", i.e. an extended Higgs mechanism,
whereby infinitely many higher spin gauge fields eat lower spin fields
and become massive, the simplest (non generic) instance being
represented by the long Konishi multiplet. Notice that these Goldstone
particles are not included in most of the studied HS theories
\cite{vas4,vas5,sezsun4,sezsun5} that are restricted to the `massless'
sector that can only capture physics at $\lambda=0$.  Infinitely many
Regge trajectories are needed that correspond to massive HS multiplets
whose interaction could still be tightly constrained by HS symmetry, as 
witnessed
by the simplicity of the anomalous dimension of HS currents, that calls
for a number theoretic origin.  Consistent coupling to the dilaton
should determine the mass shifts that are holographically dual to the
anomalous dimensions.

Although our results are rather encouraging, we expect 
the need of a refinement of our mass formula, even at $\lambda=0$, 
for states belonging to higher Regge trajectories dual to higher twist operators.
In general operators with more than two impurities are rather poorly
understood, except possibly for those with largest possible anomalous
dimension \cite{mz}. A very powerful tool for a systematic analysis 
at non-vanishing coupling is provided by the integrable
spin chain proposed in \cite{mz} and further exploited in \cite{bks},
where the spectrum of the dilatation operator \cite{bkps} for two-impurity 
operators in the BMN limit \cite{bmn} is shown to precisely match
the spectrum of the light-cone Hamilatonian in the pp-wave geometry \cite{bmn}. 
It would be nice to extend these results to ``more than two" scalar and
to fermionic impurities, the latter presumably in terms of a
supersymmetric spin chain. We believe the results presented here motivate
further investigation, possibly in relation to other
approaches (bits, tensionless strings, 
etc.)~\cite{bits,lind,lindsun,zeroten,lindzab}.

\section*{Acknowledgements}
We would like to thank N. Beisert, S. Kovacs, A. Petkou, A. Sagnotti,
E.  Sokatchev, Ya. Stanev, P. Sundell, M. Trigiante, and M. Vasiliev,
for useful discussions.  This work was supported in part by I.N.F.N.,
by the EC programs HPRN-CT-2000-00122, HPRN-CT-2000-00131 and
HPRN-CT-2000-00148, by the INTAS contract 99-1-590, by the MURST-COFIN
contract 2001-025492 and by the NATO contract PST.CLG.978785.

\begin{appendix}

\mathon
\section{Group theory formulas}
\mathoff 

{\bf $SU(4)$ product formulas:}
\bea {\bf 6}\times[k,p,q]&=&
[k-1,p,q+1]+[k-1,p+1,q-
1]+[k,p+1,q]\label{qs06}\\
                &&+[k+1,p,q-1]+[k+1,p-1,q+1]+[k,p-1,q]\;,\nn\\
{\bf 4_s}\times[k,p,q]&=&[k-1,p+1,q]+[k,p-1,q+1]+[k,p,q-
1]+[k+1,p,q]\;,\nn\\
{\bf
4_c}\times[k,p,q]&=&[k+1,p-1,q]+[k,p+1,q-1]+[k,p,q+1]+[k-1,p,q]\;.\nn
\eea
Dimension formula:
\bea
d([k,p,q])&=&
\ft{1}{12}(k\pls p\pls q\pls3)(k\pls p\pls2)
(p\pls q\pls2)(k\pls1)(p\pls1)(q\pls1)
\;.
\label{dimSU4}
\eea

\mathon
\section{$SO(9)$ content of the flat space string spectrum}
\mathoff

\label{app:vac}

\bea {\rm vac}_1 &=& [0, 0, 0, 0] ~=~ {\bf 1}
\nn\\
{\rm vac}_2 &=& [1, 0, 0, 0] ~=~ {\bf 9}
\nn\\
{\rm vac}_3 &=&      [0, 0, 0, 1]+[2, 0, 0, 0] ~=~
    {\bf 44}+{\bf 16}
\nn\\
{\rm vac}_4 &=&
    [0, 1, 0, 0]
    +[1, 0, 0, 0]
    +[1, 0, 0, 1]
    +[3, 0, 0, 0]
\nn\\ &=&
    {\bf 36}
    +{\bf 9}
    +{\bf 128}
    +{\bf 156}
\nn\\
{\rm vac}_5 &=&
     [0, 0, 0, 0]
    +[0, 0, 0, 1]
    +[0, 0, 1, 0]
    +[0, 1, 0, 0]
    +[1, 0, 0, 1]
    \nonumber\\ && {}
    +[1, 1, 0, 0]
    +[2, 0, 0, 0]
    +[2, 0, 0, 1]
    +[4, 0, 0, 0]
\nn\\ &=&
    {\bf 1}
    +{\bf 16}
    +{\bf 84}
    +{\bf 36}
    +{\bf 128}
    \nonumber\\ && {}
    +{\bf 231}
    +{\bf 44}
    +{\bf 576}
    +{\bf 450}
\nn\\
{\rm vac}_6 &=&
    2 \!\cdot\! [0, 0, 0, 1]
    +[0, 0, 0, 2]
    +[0, 1, 0, 0]
    +[0, 1, 0, 1]
    +2 \!\cdot\! [1, 0, 0, 0]
    \nonumber\\ && {}
    +2 \!\cdot\! [1, 0, 0, 1]
    +[1, 0, 1, 0]
    +2 \!\cdot\! [1, 1, 0, 0]
    +[2, 0, 0, 0]
    +[2, 0, 0, 1]
    \nonumber\\ && {}
    +[2, 1, 0, 0]
    +[3, 0, 0, 0]
    +[3, 0, 0, 1]
    +[5, 0, 0, 0]
\nn\\ &=&
    2 \!\cdot\! {\bf 16}
    +{\bf 126}
    +{\bf 36}
    +{\bf 432}
    +2 \!\cdot\! {\bf 9}
    \nonumber\\ && {}
    +2 \!\cdot\! {\bf 128}
    +{\bf 594}
    +2 \!\cdot\! {\bf 231}
    +{\bf 44}
    +{\bf 576}
    \nonumber\\ && {}
    +{\bf 910}
    +{\bf 156}
    +{\bf 1920}
    +{\bf 1122}
\nn\\
{\rm vac}_7 &=&
    2 \!\cdot\! [0, 0, 0, 0]
    +2 \!\cdot\! [0, 0, 0, 1]
    +[0, 0, 0, 2]
    +3 \!\cdot\! [0, 0, 1, 0]
    +2 \!\cdot\! [0, 1, 0, 0]
    \nonumber\\ && {}
    +2 \!\cdot\! [0, 1, 0, 1]
    +[0, 2, 0, 0]
    +2 \!\cdot\! [1, 0, 0, 0]
    +4 \!\cdot\! [1, 0, 0, 1]
    +[1, 0, 0, 2]
    \nonumber\\ && {}
    +[1, 0, 1, 0]
    +2 \!\cdot\! [1, 1, 0, 0]
    +[1, 1, 0, 1]
    +3 \!\cdot\! [2, 0, 0, 0]
    +3 \!\cdot\! [2, 0, 0, 1]
    \nonumber\\ && {}
    +[2, 0, 1, 0]
    +2 \!\cdot\! [2, 1, 0, 0]
    +[3, 0, 0, 0]
    +[3, 0, 0, 1]
    +[3, 1, 0, 0]
    \nonumber\\ && {}
    +[4, 0, 0, 0]
    +[4, 0, 0, 1]
    +[6, 0, 0, 0]
\nn\\ &=&
    2 \!\cdot\! {\bf 1}
    +2 \!\cdot\! {\bf 16}
    +{\bf 126}
    +3 \!\cdot\! {\bf 84}
    +2 \!\cdot\! {\bf 36}
    \nonumber\\ && {}
    +2 \!\cdot\! {\bf 432}
    +{\bf 495}
    +2 \!\cdot\! {\bf 9}
    +4 \!\cdot\! {\bf 128}
    +{\bf 924}
    \nonumber\\ && {}
    +{\bf 594}
    +2 \!\cdot\! {\bf 231}
    +{\bf 2560}
    +3 \!\cdot\! {\bf 44}
    +3 \!\cdot\! {\bf 576}
    \nonumber\\ && {}
    +{\bf 2457}
    +2 \!\cdot\! {\bf 910}
    +{\bf 156}
    +{\bf 1920}
    +{\bf 2772}
    \nonumber\\ && {}
    +{\bf 450}
    +{\bf 5280}
    +{\bf 2508}
\eea

\mathon
\section{Decomposition of long multiplets}
\mathoff

\label{app:shortenings}

Let us briefly describe the ${\cal N}=4$ superconformal multiplet
structure following~\cite{dolosb}. The long supermultiplet \bea {\cal
A}^{\Delta_0}_{[k,p,q](j,\Bj)} &\equiv& T_1\times
[k,p,q]_{(j,\Bj)} \;, \la{AinT} \eea 
is obtained by the unconstrained
action of all sixteen supercharges on the state
$[k,p,q]_{(j,\Bj)}$ of lowest conformal dimension $\Delta_0$. The
precise representation content may be found from evaluating the
tensor product~\Ref{AinT}, or explicitly by using the
Racah-Speiser algorithm as
\bea
{\cal
A}^{\Delta_0}_{[k,p,q](j,\Bj)}
 &=&
\sum_{\epsilon_{i\alpha},\bar{\epsilon}^i_{\dot{\alpha}}\in
\{0,1\}} \left[ (k,p,q)_{(j,\Bj)}+\epsilon_{i\alpha}q^{i
\alpha} + \bar{\epsilon}^i_{\dot{\alpha}} \bar{q}^{\dot{\alpha}}_i
\right] \;,
 \label{susy}
 \eea
with the sum running over the $2^{16}$
combinations of the $16$ weights $q^{i \pm}$, $\bar{q}^{\pm}_i$,
$i=1, \dots 4$ of the supersymmetry charges
\bea
q^{1\pm}&=&[1,0,0]_{(\pm \frac12,0)}\;, \quad\quad
~~~q^{2\pm}=[-1,1,0]_{(\pm \frac12,0)}\;,\nn\\
q^{3\pm}&=&[0,-1,1]_{(\pm \frac12,0)} \;,\quad\quad
q^{4\pm}=[0,0,-1]_{(\pm\frac12,0)} \;,\nn\\
\bar{q}_1^{\pm}&=&[0,0,1]_{(0,\pm\frac12)} \;,\quad\quad
~~~\bar{q}_2^{\pm}=[0,1,-1]_{(0,\pm\frac12)}\;,\nn\\
\bar{q}_3^{\pm}&=&[1,-1,0]_{(0,\pm\frac12)}\;,\quad\quad
 \bar{q}_4^{\pm}=[-1,0,0]_{(0,\pm\frac12)}\;.
\label{qs}
\eea
Every $q$, $\bar{q}$ raises the conformal dimension by $1/2$. In order
to make sense out of \Ref{susy} also for
small values of $k, p, q, j, \Bj$, we need to count
representations with negative weights according to the
identifications
\bea
[k,p,q]_{(j,\Bj)}&=&-[-k\mis2,p\pls k\pls
1,q]_{(j,\Bj)} ~=~ -[k,p\pls q\pls 1,-q\mis2]_{(j,\Bj)} \non &=&
-[k\pls p\pls 1,-p\mis2,q\pls p\pls 1]_{(j,\Bj)} \;, \non &=&
-[k,p,q]_{(-j-1,\Bj)} ~=~ -[k,p,q]_{(j,-\Bj-1)}
 \;.
\la{negw}
\eea
In particular, this implies that
$[k,p,q]_{(j,\Bj)}$ is counted as zero whenever any of the weights
$k$, $p$, $q$ equals~$-1$ or one of the spins $j$, $\Bj$
equals $-\ft12$.

Short and semi-short multiplets are obtained by acting on the
ground state only with a limited number of supercharges, i.e.\ the
sum in \Ref{susy} is restricted to a sum over partial subsets of
weights $q^{i \pm}$'s, $\bar{q}^{\pm}_i$'s. This requires certain
conditions on the ground state $[k,p,q]_{(j,\Bj)}$ and its
conformal dimension $\Delta_0$. In table~\ref{boundtype} we have
collected the possible subsets of $q^{i \pm}$'s together with the
corresponding restrictions on the ground state. Analogous
conditions hold for the $\bar{q}_i^{\pm}$'s. The general multiplet
is obtained by combining a subset of $q^{i \pm}$'s with a subset
of $\bar{q}_i^{\pm}$'s provided the ground state
$[k,p,q]_{(j,\Bj)}$ satisfies both constraints. Corresponding
to the type of shortening and the fraction of preserved
supersymmetry, we denote them by ${\cal B}{\cal B}^{s,\bar s}$,
${\cal B}{\cal C}^{s,\bar t}$, etc.

\begin{table}[bt]
\centering
\begin{tabular}{cclll}
type & fraction of susy & HWS & with conformal dimension
$\Delta_0$ & subset of weights \cr\hline
 ${\cal B}$ &$s=\ft14$ & $[k,p,q]_{(0,\Bj)}$&
$\ft12(3k+2p+q)$ & $q_{\cal B}=q_{2,3,4}^{\pm}$ \cr   ${\cal B}$ &
$s=\ft12$ &  $[0,p,q]_{(0,\Bj)}$& $\ft12(2p+q)$ &$q_{\cal
B}=q_{3,4}^{\pm}$\cr   ${\cal B}$ &  $s=\ft34$ &
$[0,0,q]_{(0,\Bj)}$& $\ft12 q$ &$q_{\cal B}=q_{4}^{\pm}$ \cr
${\cal B}$ &$s=1$ &  $[0,0,0]_{(0,\Bj)}$& $0$ &$q_{\cal
B}=\emptyset$ \cr   ${\cal C}$ &$t=\ft14$ & $[k,p,q]_{(j,\Bj)}$&
$2+2j+\ft12(3k+2p+q)$ & $q_{\cal C}=q_1^+,q_{2,3,4}^{\pm}$ \cr
${\cal C}$  & $t=\ft12$ & $[0,p,q]_{(j,\Bj)}$&
$2+2j+\ft12(2p+q)$ &$q_{\cal C}=q_{1,2}^+,q_{3,4}^{\pm}$\cr   ${\cal
C}$ &
 $t=\ft34$ &  $[0,0,q]_{(j,\Bj)}$& $2+2j+\ft12 q$
&$q_{\cal C}=q_{1,2,3}^+,q_{4}^{\pm}$ \cr   ${\cal C}$ &$t=1$ &
$[0,0,0]_{(j,\Bj)}$& $2+2j$ &$q_{\cal C}=q_{1,2,3,4}^+$ \cr
\end{tabular}
\caption{\small BPS shortening (${\cal B}$) and semi-shortening
(${\cal C}$) conditions among the $q^{i \pm}$. Similar relations
hold for the weights $\bar{q}_i^{\pm}$ of the conjugate
supercharges with $[k,p,q]_{(j,\Bj)}\leftrightarrow
[q,p,k]_{(\Bj,j)}$.} \label{boundtype}
\end{table}

E.g.\ the BPS multiplets are for instance ${\cal B}{\cal
B}^{\frac12,\frac12}_{[0,n,0](0,0)}$, and ${\cal B}{\cal
B}^{\frac14,\frac14}_{[q,p,q](0,0)}$ with lowest conformal dimension
$\Delta_0 = n, p+2q$, respectively. The former are realized in the KK
tower of the $\ell=0$ supergravity sector while the latter appear
typically in the decomposition (\ref{decomposition}) of the so called
``BMN multiplets" ${\cal A}_{[q\mis2,p,q\mis2]}^{p+2q-2}$.  The KK
supergravity multiplets are given in table~\ref{Tsugra} below with
total dimension
\bea
\dim {\cal B}{\cal
B}^{\frac12,\frac12}_{[0,n,0](0,0)} &=& 2^8\, {\rm dim}\,[0,n-2,0]\;.
\eea
In particular, for $n=2$ this is the massless supergravity multiplet
of dimension 256. Strictly speaking, massless states transform only
under the diagonal $SU(2)$ subgroup defining the helicity
rotations. We will indicate this by a superscript `$^*$', e.g.\
$(j,\Bj)=(\ft12,\ft12)^*$ denotes a massless state of helicity~1. This
is important for the correct counting of degrees of freedom. Another
interesting multiplet it the semishort massless multiplet ${\cal
C}{\cal C}^{1,1}_{[0,0,0](\ell-1,\ell-1)^*}$ with lowest conformal
dimension $\Delta_0 = 2\ell$, which we give in table~\ref{C11}. Its
total dimension is given by $256\,(4\ell\pls1)$.

Consider now the generic long multiplet ${\cal
A}^{\Delta_0}_{[k,p,q](j,\Bj)}$. Unitarity requires the
bounds~\cite{dp}
\bea
\Delta_0 &\ge& 2+2j+\ft12(3k+2p+q) \;,\qquad \Delta_0 ~\ge~
2+2\Bj+\ft12(3q+2p+k)
\;, \la{unitarity}
\eea
for the lowest conformal dimension $\Delta_0$ (unless $j=0$ or $\Bj=0$
and the shortening is of BPS type, in which case also
$\Delta=\ft12(3k+2p+q)$ and $\Delta=\ft12(3q+2p+k)$, respectively, are
allowed). Simultaneous saturation of the two bounds~\Ref{unitarity}
implies $k-q\equiv2(\Bj-j)$ and corresponds to a ${\cal C}{\cal C}$
type shortening of the multiplet. More precisely, the long multiplet
then decomposes into~\cite{dolosb}
\bea {\cal A}^{2+2j+\frac12(3k+2p+q)}_{\left[k,p,q
\right](j,\Bj)}\big|_{k-q=2(\bar{j}-j)} &\longrightarrow& {\cal C}{\cal
C}^{\frac14,\frac14}_{\left[k,p,q\right](j,\Bj)}+ {\cal
C}{\cal
C}^{\frac14,\frac14}_{\left[k+1,p,q\right](j-\frac12,\Bj)}+
\non &&{} + {\cal C}{\cal
C}^{\frac14,\frac14}_{\left[k,p,q+1\right](j,\Bj-\frac12)}+
{\cal C}{\cal
C}^{\frac14,\frac14}_{\left[k+1,p,q+1\right](j-\frac12,\Bj-\frac12)}
\;.
\eea
For particular values of $k,p,q, j, \Bj$, some of the appearing
semi-short multiplets preserve more supersymmetry, and the
decomposition becomes
\footnote{ We note that there is a misprint in the dimension formula
  (5.49) of \cite{dolosb} which should read \ben {\rm dim} \bar{\cal
    D}^{\frac34,\frac14}_{[0,0,q](j,0)} = 2^{11} d(0,0,q-1)({\cal J} +
  {3\over 2}) - 256 [(2\hat{q}^2 - 1) ({\cal J}\underline{+1})
    +\hat{q}]\;, \een where $\hat{q}= q+1$ ${\cal J} = 2j+1$, and
  $d(k,p,q)$ from \Ref{dimSU4}. The missing term in \cite{dolosb} is
  underlined in the above formula. We checked the correct formula in
  the decomposition \Ref{decomposition}.}
\bea {\cal
A}^{2+2j+p}_{\left[0p0\right](j,j)}&\longrightarrow & {\cal
C}{\cal C}^{{1\over 2},{1\over 2}}_{\left[0,p,0\right](j,j)}+
{\cal C}{\cal C}^{{1\over 4},{1\over
2}}_{\left[1,p,0\right](j-\frac12,j)}+ {\cal C}{\cal C}^{{1\over
2},{1\over 4}}_{\left[0,p,1\right](j,j-\frac12)}+ {\cal C}{\cal
C}^{{1\over 4},{1\over
4}}_{\left[1,p,1\right](j-\frac12,j-\frac12)} \;, \non {\cal
A}^{2+2j}_{\left[000\right](j,j)^*}&\longrightarrow & {\cal
C}{\cal C}^{{1},{1}}_{\left[0,0,0\right](j,j)^*}+ {\cal C}{\cal
C}^{{1\over 4},{3\over 4}}_{\left[1,0,0\right](j-\frac12,j)}+
{\cal C}{\cal C}^{{3\over 4},{1\over
4}}_{\left[0,0,1\right](j,j-\frac12)}+ {\cal C}{\cal C}^{{1\over
4},{1\over 4}}_{\left[1,0,1\right](j-\frac12,j-\frac12)} \;, \non
{\cal A}^{2+p+2q}_{\left[q,p,q\right](0,0)}&\longrightarrow &
{\cal C}{\cal C}^{{1\over 4},{1\over
4}}_{\left[q,p,q\right](0,0)}+ {\cal B}{\cal C}^{{1\over
4},{1\over 4}}_{\left[q+2,p,q\right](0,0)}+ {\cal C}{\cal
B}^{{1\over 4},{1\over 4}}_{\left[q,p,q+2\right](0,0)}+ {\cal
B}{\cal B}^{{1\over 4},{1\over 4}}_{\left[q+2,p,q+2\right](0,0)}
\;, \non {\cal A}^{2+p}_{\left[0p0\right](0,0)}&\longrightarrow &
{\cal C}{\cal C}^{{1\over 2},{1\over
2}}_{\left[0,p,0\right](0,0)}+ {\cal B}{\cal C}^{{1\over
4},{1\over 2}}_{\left[2,p,0\right](0,0)}+ {\cal C}{\cal
B}^{{1\over 2},{1\over 4}}_{\left[0,p,2\right](0,0)}+ {\cal
B}{\cal B}^{{1\over 4},{1\over 4}}_{\left[2,p,2\right](0,0)} \;,
\non {\cal A}^{2}_{\left[000\right](0,0)^*}&\longrightarrow &
{\cal
C}{\cal C}^{{1},{1}}_{\left[000\right](0,0)^*}+ {\cal B}{\cal
C}^{{1\over 4},{3\over 4}}_{\left[200\right](0,0)}+ {\cal C}{\cal
B}^{{3\over 4},{1\over 4}}_{\left[002\right](0,0)}+ {\cal B}{\cal
B}^{{1\over 4},{1\over 4}}_{\left[202\right](0,0)} \;.
\label{decomposition} \eea

\end{appendix}

\newpage

\begin{table}[htb]
     \centering
{\footnotesize
\begin{tabular}{|c|l|} \hline
     $\Delta$ &\\ \hline
     $n$ & $[0,n,0]_{(0,0)}$ \\[.5ex]
     $n\pls\ft12$ & $[1,n\mis1,0]_{(0,\frac12)}+
[0,n\mis1,1]_{(\frac12,0)}$ \\[.5ex]
     $n\pls1$ & $[0,n\mis2,2]_{(0,0)}+[2,n\mis2,0]_{(0,0)}
+[0,n\mis1,0]_{(0,1)}+[0,n\mis1,0]_{(1,0)}
+[1,n\mis2,1]_{(\frac12,\frac12)}$ \\[.5ex]
     $n\pls\ft32$ & $[1,n\mis2,0]_{(0,\frac12)}+[1,n\mis3,2]_{(0,\frac12)}
+[0,n\mis2,1]_{(\frac12,0)}+[2,n\mis3,1]_{(\frac12,0)}+
[0,n\mis2,1]_{(\frac12,1)}+[1,n\mis2,0]_{(1,\frac12)}$ \\[.5ex]
     $n\pls2$ & $2\!\cdot\![0,n\mis2,0]_{(0,0)}+[2,n\mis4,2]_{(0,0)}
+[0,n\mis3,2]_{(0,1)}+[2,n\mis3,0]_{(1,0)}
+2\!\cdot\![1,n\mis3,1]_{(\frac12,\frac12)}+[0,n\mis2,0]_{(1,1)}$
\\[.5ex]
$n\pls\ft52$ &
$[1,n\mis3,0]_{(0,\frac12)}+[1,n\mis4,2]_{(0,\frac12)}+
[0,n\mis3,1]_{(\frac12,0)}+[2,n\mis4,1]_{(\frac12,0)}
+[0,n\mis3,1]_{(\frac12,1)}+[1,n\mis3,0]_{(1,\frac12)}$ \\[.5ex]
     $n\pls3$ & $[0,n\mis4,2]_{(0,0)}+[2,n\mis4,0]_{(0,0)}
+[0,n\mis3,0]_{(0,1)}+[0,n\mis3,0]_{(1,0)}
+[1,n\mis4,1]_{(\frac12,\frac12)}$ \\[.5ex]
     $n\pls\ft{7}2$ & $[1,n\mis4,0]_{(0,\frac12)}+
[0,n\mis4,1]_{(\frac12,0)}$ \\[.5ex]
     $n\pls4$ & $[0,n\mis4,0]_{(0,0)}$ \\
\hline
\end{tabular}
}
\caption{{\small BPS multiplet
$\CB\CB^{\frac12,\frac12}_{[0,n,0](0,0)}$ with $n\geq 2$. 
Negative weights are 
understood
to be omitted. The total number of states is $d(n)= 2^8\cdot {\rm
dim}\,[0,n-2,0]$.
}}
\label{Tsugra}
\end{table}

\begin{table}[htb]
     \centering
{\footnotesize
\begin{tabular}{|c|l|} \hline
     $\Delta$ &\\ \hline
     $2\ell$ & $[0,0,0]_{(\ell-1,\ell-1)}$ \\[1ex]
     $2\ell\pls\ft12$ & $[0,0,1]_{(\ell-1,\ell-\frac12)}+
[1,0,0]_{(\ell-\frac12,\ell-1)}$ \\[1ex]
     $2\ell\pls1$ & $[0,1,0]_{(\ell-
1,\ell)}+[0,1,0]_{(\ell,\ell-1)}
+[0,0,0]_{(\ell-\frac12,\ell-\frac12)} +[1,0,1]_{(\ell-
\frac12,\ell-\frac12)}$
\\[1ex]
     $2\ell\pls\ft32$ &
$[1,0,0]_{(\ell-1,\ell+\frac12)+(\ell,\ell-\frac12)}+
[0,0,1]_{(\ell+\frac12,\ell-
1)+(\ell-\frac12,\ell)}+[1,1,0]_{(\ell-
\frac12,\ell)}+[0,1,1]_{(\ell,\ell-
\frac12)}$ \\[1ex]
     $2\ell\pls2$ &
$[0,0,0]_{(\ell-1,\ell+1)+(\ell,\ell)+(\ell+1,\ell-
1)}+[0,1,0]_{(\ell-\frac12,\ell+\frac12)+(\ell+\frac12,\ell-
\frac12)}+[1,0,1]_{(\ell,\ell)}+[0,2,0]_{(\ell,\ell)} $
\non[.5ex]
&$+[2,0,0]_{(\ell-
\frac12,\ell+\frac12)}+[0,0,2]_{(\ell+\frac12,\ell-
\frac12)}$
\\[1ex]
$2\ell\pls\ft52$ &
$[1,0,0]_{(\ell-
\frac12,\ell+1)+(\ell+\frac12,\ell)}+[0,0,1]_{(\ell+1,\ell-
\frac12)+(\ell,\ell+\frac12)}+[1,1,0]_{(\ell,\ell+\frac12)}+
[0,1,1]_{(\ell+\frac12,\ell)}$
\\[1ex]
      $2\ell\pls3$ & 
$[0,1,0]_{(\ell,\ell+1)}+[0,1,0]_{(\ell+1,\ell)}
+[0,0,0]_{(\ell+\frac12,\ell+\frac12)} +[1,0,1]_{(\ell+
\frac12,\ell+\frac12)}$
\\[1ex]
     $2\ell\pls\ft{7}2$ & 
      $[0,0,1]_{(\ell+\ft12,\ell+1)}+
[1,0,0]_{(\ell+1,\ell+\ft12)}$ \\[1ex]
     $2\ell\pls4$ & $[0,0,0]_{(\ell+1,\ell+1)}$ \\
\hline
\end{tabular}
}
\caption{{\small Massless multiplet
$\CC\CC^{1,1}_{[0,0,0](\ell-1,\ell-1)^*}$. 
The number of physical states is $d(\ell)=
2^8\cdot(4\ell+1).$}}
\label{C11}
\end{table}

\begin{table}[htb] 
\centering
{\footnotesize
\begin{tabular}{|c|l|} \hline
     $\Delta$ & \\ \hline\hline
     2& $
     [0,0,0]_{(0,0)}
     $ \\ \hline $
     \frac{5}{2}$& $
     [0,0,1]_{(0,\frac{1}{2})}
     +[1,0,0]_{(\frac{1}{2},0)}
     $ \\ \hline $
     3$& $
     [0,0,0]_{(\frac{1}{2},\frac{1}{2})}
     +[0,0,2]_{(0,0)}
     +[0,1,0]_{(0,1)+(1,0)}
     +[1,0,1]_{(\frac{1}{2},\frac{1}{2})}
     +[2,0,0]_{(0,0)}
     $ \\ \hline $
     \frac{7}{2}$& $
     [0,0,1]_{(\frac{1}{2},0)+(\frac{1}{2},1)+(\frac{3}{2},0)}
     +[0,1,1]_{(0,\frac{1}{2})+(1,\frac{1}{2})}
     +[1,0,0]_{(0,\frac{1}{2})+(0,\frac{3}{2})+(1,\frac{1}{2})}
     +[1,0,2]_{(\frac{1}{2},0)}
     $\\& $
     +[1,1,0]_{(\frac{1}{2},0)+(\frac{1}{2},1)}
     +[2,0,1]_{(0,\frac{1}{2})}
     $ \\ \hline $
     4$& $
     [0,0,0]_{(0,0)+(0,2)+(1,1)+(2,0)}
     +[0,0,2]_{(\frac{1}{2},\frac{1}{2})+(\frac{3}{2},\frac{1}{2})}
     +[0,1,0]_{2(\frac{1}{2},\frac{1}{2})+(\frac{1}{2},\frac{3}{2})+(\frac{3}{2},\frac{1}{2})}
     +[2,0,2]_{(0,0)}
     +[2,1,0]_{(0,1)}
     $\\& $
     +[0,1,2]_{(1,0)}
     +[0,2,0]_{2(0,0)+(1,1)}
     +[1,0,1]_{(0,0)+2(0,1)+2(1,0)+(1,1)}
     +[1,1,1]_{2(\frac{1}{2},\frac{1}{2})}
     +[2,0,0]_{(\frac{1}{2},\frac{1}{2})+(\frac{1}{2},\frac{3}{2})}
     $ \\ \hline $
     \frac{9}{2}$& $
     [0,0,1]_{(0,\frac{1}{2})+(0,\frac{3}{2})+2(1,\frac{1}{2})+(1,\frac{3}{2})+(2,\frac{1}{2})}
     +[0,0,3]_{(\frac{3}{2},0)}
     +[0,1,1]_{3(\frac{1}{2},0)+2(\frac{1}{2},1)+(\frac{3}{2},0)+(\frac{3}{2},1)}
     +[0,2,1]_{(0,\frac{1}{2})+(1,\frac{1}{2})}
     $\\& $
     +[1,0,0]_{(\frac{1}{2},0)+2(\frac{1}{2},1)+(\frac{1}{2},2)+(\frac{3}{2},0)+(\frac{3}{2},1)}
     +[1,0,2]_{(0,\frac{1}{2})+2(1,\frac{1}{2})}
     +[1,1,0]_{3(0,\frac{1}{2})+(0,\frac{3}{2})+2(1,\frac{1}{2})+(1,\frac{3}{2})}
     $\\& $
     +[1,1,2]_{(\frac{1}{2},0)}
     +[1,2,0]_{(\frac{1}{2},0)+(\frac{1}{2},1)}
     +[2,0,1]_{(\frac{1}{2},0)+2(\frac{1}{2},1)}
     +[2,1,1]_{(0,\frac{1}{2})}
     +[3,0,0]_{(0,\frac{3}{2})}
     $ \\ \hline $
     5$& $
     [0,0,0]_{(\frac{1}{2},\frac{1}{2})+(\frac{1}{2},\frac{3}{2})+(\frac{3}{2},\frac{1}{2})+(\frac{3}{2},\frac{3}{2})}
     +[0,0,2]_{2(0,0)+(1,0)+2(1,1)+(2,0)}
     +[0,1,0]_{3(0,1)+3(1,0)+2(1,1)+(1,2)+(2,1)}
     $\\& $
     +[0,1,2]_{2(\frac{1}{2},\frac{1}{2})+(\frac{3}{2},\frac{1}{2})}
     +[0,2,0]_{3(\frac{1}{2},\frac{1}{2})+(\frac{1}{2},\frac{3}{2})+(\frac{3}{2},\frac{1}{2})}
     +[0,2,2]_{(0,0)}
     +[0,3,0]_{(0,1)+(1,0)}
     $\\& $
     +[1,0,3]_{(1,0)}
     +[1,1,1]_{2(0,0)+2(0,1)+2(1,0)+2(1,1)}
     +[1,2,1]_{(\frac{1}{2},\frac{1}{2})}
     +[2,0,0]_{2(0,0)+(0,1)+(0,2)+2(1,1)}
     $\\& $
     +[2,0,2]_{(\frac{1}{2},\frac{1}{2})}
     +[2,1,0]_{2(\frac{1}{2},\frac{1}{2})+(\frac{1}{2},\frac{3}{2})}
     +[2,2,0]_{(0,0)}
     +[3,0,1]_{(0,1)}
     +[1,0,1]_{4(\frac{1}{2},\frac{1}{2})+2(\frac{1}{2},\frac{3}{2})+2(\frac{3}{2},\frac{1}{2})+(\frac{3}{2},\frac{3}{2})}
     $ \\ \hline $
     \frac{11}{2}$& $
     [0,0,1]_{2(\frac{1}{2},0)+3(\frac{1}{2},1)+(\frac{3}{2},0)+2(\frac{3}{2},1)+(\frac{3}{2},2)}
     +[0,0,3]_{(0,\frac{1}{2})+(1,\frac{1}{2})}
     +[0,1,1]_{3(0,\frac{1}{2})+(0,\frac{3}{2})+4(1,\frac{1}{2})+2(1,\frac{3}{2})+(2,\frac{1}{2})}
     $\\& $
     +[0,1,3]_{(\frac{1}{2},0)}
     +[0,2,1]_{2(\frac{1}{2},0)+2(\frac{1}{2},1)+(\frac{3}{2},0)}
     +[0,3,1]_{(0,\frac{1}{2})}
     +[1,0,0]_{2(0,\frac{1}{2})+(0,\frac{3}{2})+3(1,\frac{1}{2})+2(1,\frac{3}{2})+(2,\frac{3}{2})}
     $\\& $
     +[1,0,2]_{2(\frac{1}{2},0)+2(\frac{1}{2},1)+(\frac{3}{2},0)+(\frac{3}{2},1)}
     +[1,1,0]_{3(\frac{1}{2},0)+4(\frac{1}{2},1)+(\frac{1}{2},2)+(\frac{3}{2},0)+2(\frac{3}{2},1)}
     +[1,1,2]_{(0,\frac{1}{2})+(1,\frac{1}{2})}
     $\\& $
     +[1,2,0]_{2(0,\frac{1}{2})+(0,\frac{3}{2})+2(1,\frac{1}{2})}
     +[1,3,0]_{(\frac{1}{2},0)}
     +[2,0,1]_{2(0,\frac{1}{2})+(0,\frac{3}{2})+2(1,\frac{1}{2})+(1,\frac{3}{2})}
     +[2,1,1]_{(\frac{1}{2},0)+(\frac{1}{2},1)}
     $\\& $
     +[3,0,0]_{(\frac{1}{2},0)+(\frac{1}{2},1)}
     +[3,1,0]_{(0,\frac{1}{2})}
     $ \\ \hline $
     6$& $
     [0,0,0]_{3(0,0)+3(1,1)+(2,2)}
     +[0,0,2]_{3(\frac{1}{2},\frac{1}{2})+(\frac{1}{2},\frac{3}{2})+(\frac{3}{2},\frac{1}{2})+(\frac{3}{2},\frac{3}{2})}
     +[0,1,0]_{4(\frac{1}{2},\frac{1}{2})+2(\frac{1}{2},\frac{3}{2})+2(\frac{3}{2},\frac{1}{2})+2(\frac{3}{2},\frac{3}{2})}
     $\\& $
     +[0,1,2]_{(0,0)+2(0,1)+2(1,0)+(1,1)}
     +[0,2,0]_{3(0,0)+(0,1)+(0,2)+(1,0)+3(1,1)+(2,0)}
     +[0,2,2]_{(\frac{1}{2},\frac{1}{2})}
     $\\& $
     +[0,3,0]_{2(\frac{1}{2},\frac{1}{2})}
     +[0,4,0]_{(0,0)}
     +[1,0,1]_{(0,0)+3(0,1)+3(1,0)+4(1,1)+(1,2)+(2,1)}
     +[1,0,3]_{(\frac{1}{2},\frac{1}{2})}
     +[0,0,4]_{(0,0)}
     $\\& $
     +[1,1,1]_{4(\frac{1}{2},\frac{1}{2})+2(\frac{1}{2},\frac{3}{2})+2(\frac{3}{2},\frac{1}{2})}
     +[1,2,1]_{(0,0)+(0,1)+(1,0)}
     +[2,0,0]_{3(\frac{1}{2},\frac{1}{2})+(\frac{1}{2},\frac{3}{2})+(\frac{3}{2},\frac{1}{2})+(\frac{3}{2},\frac{3}{2})}
     $\\& $
     +[2,0,2]_{(0,0)+(1,1)}
     +[2,1,0]_{(0,0)+2(0,1)+2(1,0)+(1,1)}
     +[2,2,0]_{(\frac{1}{2},\frac{1}{2})}
     +[3,0,1]_{(\frac{1}{2},\frac{1}{2})}
     +[4,0,0]_{(0,0)}
     $ \\ \hline $
     \frac{13}{2}$& $
     [0,0,1]_{2(0,\frac{1}{2})+(0,\frac{3}{2})+3(1,\frac{1}{2})+2(1,\frac{3}{2})+(2,\frac{3}{2})}
     +[0,0,3]_{(\frac{1}{2},0)+(\frac{1}{2},1)}
     +[0,1,1]_{3(\frac{1}{2},0)+4(\frac{1}{2},1)+(\frac{1}{2},2)+(\frac{3}{2},0)+2(\frac{3}{2},1)}
     $\\& $
     +[0,1,3]_{(0,\frac{1}{2})}
     +[0,2,1]_{2(0,\frac{1}{2})+(0,\frac{3}{2})+2(1,\frac{1}{2})}
     +[0,3,1]_{(\frac{1}{2},0)}
     +[1,0,0]_{2(\frac{1}{2},0)+3(\frac{1}{2},1)+(\frac{3}{2},0)+2(\frac{3}{2},1)+(\frac{3}{2},2)}
     $\\& $
     +[1,0,2]_{2(0,\frac{1}{2})+(0,\frac{3}{2})+2(1,\frac{1}{2})+(1,\frac{3}{2})}
     +[1,1,0]_{3(0,\frac{1}{2})+(0,\frac{3}{2})+4(1,\frac{1}{2})+2(1,\frac{3}{2})+(2,\frac{1}{2})}
     +[1,1,2]_{(\frac{1}{2},0)+(\frac{1}{2},1)}
     $\\& $
     +[1,2,0]_{2(\frac{1}{2},0)+2(\frac{1}{2},1)+(\frac{3}{2},0)}
     +[1,3,0]_{(0,\frac{1}{2})}
     +[2,0,1]_{2(\frac{1}{2},0)+2(\frac{1}{2},1)+(\frac{3}{2},0)+(\frac{3}{2},1)}
     +[2,1,1]_{(0,\frac{1}{2})+(1,\frac{1}{2})}
     $\\& $
     +[3,0,0]_{(0,\frac{1}{2})+(1,\frac{1}{2})}
     +[3,1,0]_{(\frac{1}{2},0)}
     $ \\ \hline $
     7$& $
     [0,0,0]_{(\frac{1}{2},\frac{1}{2})+(\frac{1}{2},\frac{3}{2})+(\frac{3}{2},\frac{1}{2})+(\frac{3}{2},\frac{3}{2})}
     +[0,0,2]_{2(0,0)+(0,1)+(0,2)+2(1,1)}
     +[0,1,0]_{3(0,1)+3(1,0)+2(1,1)+(1,2)+(2,1)}
     $\\& $
     +[0,1,2]_{2(\frac{1}{2},\frac{1}{2})+(\frac{1}{2},\frac{3}{2})}
     +[0,2,0]_{3(\frac{1}{2},\frac{1}{2})+(\frac{1}{2},\frac{3}{2})+(\frac{3}{2},\frac{1}{2})}
     +[0,2,2]_{(0,0)}
     +[0,3,0]_{(0,1)+(1,0)}
     $\\& $
     +[1,0,3]_{(0,1)}
     +[1,1,1]_{2(0,0)+2(0,1)+2(1,0)+2(1,1)}
     +[1,2,1]_{(\frac{1}{2},\frac{1}{2})}
     +[2,0,0]_{2(0,0)+(1,0)+2(1,1)+(2,0)}
     $\\& $
     +[2,0,2]_{(\frac{1}{2},\frac{1}{2})}
     +[2,1,0]_{2(\frac{1}{2},\frac{1}{2})+(\frac{3}{2},\frac{1}{2})}
     +[2,2,0]_{(0,0)}
     +[3,0,1]_{(1,0)}
     +[1,0,1]_{4(\frac{1}{2},\frac{1}{2})+2(\frac{1}{2},\frac{3}{2})+2(\frac{3}{2},\frac{1}{2})+(\frac{3}{2},\frac{3}{2})}
     $ \\ \hline $
     \frac{15}{2}$& $
     [0,0,1]_{(\frac{1}{2},0)+2(\frac{1}{2},1)+(\frac{1}{2},2)+(\frac{3}{2},0)+(\frac{3}{2},1)}
     +[0,0,3]_{(0,\frac{3}{2})}
     +[0,1,1]_{3(0,\frac{1}{2})+(0,\frac{3}{2})+2(1,\frac{1}{2})+(1,\frac{3}{2})}
     +[0,2,1]_{(\frac{1}{2},0)+(\frac{1}{2},1)}
     $\\& $
     +[1,0,0]_{(0,\frac{1}{2})+(0,\frac{3}{2})+2(1,\frac{1}{2})+(1,\frac{3}{2})+(2,\frac{1}{2})}
     +[1,0,2]_{(\frac{1}{2},0)+2(\frac{1}{2},1)}
     +[1,1,0]_{3(\frac{1}{2},0)+2(\frac{1}{2},1)+(\frac{3}{2},0)+(\frac{3}{2},1)}
     $\\& $
     +[1,1,2]_{(0,\frac{1}{2})}
     +[1,2,0]_{(0,\frac{1}{2})+(1,\frac{1}{2})}
     +[2,0,1]_{(0,\frac{1}{2})+2(1,\frac{1}{2})}
     +[2,1,1]_{(\frac{1}{2},0)}
     +[3,0,0]_{(\frac{3}{2},0)}
     $ \\ \hline $
     8$& $
     [0,0,0]_{(0,0)+(0,2)+(1,1)+(2,0)}
     +[0,0,2]_{(\frac{1}{2},\frac{1}{2})+(\frac{1}{2},\frac{3}{2})}
     +[0,1,0]_{2(\frac{1}{2},\frac{1}{2})+(\frac{1}{2},\frac{3}{2})+(\frac{3}{2},\frac{1}{2})}    +[2,0,2]_{(0,0)}
     +[2,1,0]_{(1,0)}
     $\\& $
     +[0,1,2]_{(0,1)}
     +[0,2,0]_{2(0,0)+(1,1)}
     +[1,0,1]_{(0,0)+2(0,1)+2(1,0)+(1,1)}
     +[1,1,1]_{2(\frac{1}{2},\frac{1}{2})}
     +[2,0,0]_{(\frac{1}{2},\frac{1}{2})+(\frac{3}{2},\frac{1}{2})}
     $ \\ \hline $
     \frac{17}{2}$& $
     [0,0,1]_{(0,\frac{1}{2})+(0,\frac{3}{2})+(1,\frac{1}{2})}
     +[0,1,1]_{(\frac{1}{2},0)+(\frac{1}{2},1)}
     +[1,0,0]_{(\frac{1}{2},0)+(\frac{1}{2},1)+(\frac{3}{2},0)}
     +[1,0,2]_{(0,\frac{1}{2})}
     $\\& $
     +[1,1,0]_{(0,\frac{1}{2})+(1,\frac{1}{2})}
     +[2,0,1]_{(\frac{1}{2},0)}
     $ \\ \hline $
     9$& $
     [0,0,0]_{(\frac{1}{2},\frac{1}{2})}
     +[0,0,2]_{(0,0)}
     +[0,1,0]_{(0,1)+(1,0)}
     +[1,0,1]_{(\frac{1}{2},\frac{1}{2})}
     +[2,0,0]_{(0,0)}
     $ \\ \hline $
     \frac{19}{2}$ & $
     [0,0,1]_{(\frac{1}{2},0)}
     +[1,0,0]_{(0,\frac{1}{2})}
     $ \\ \hline $
     10$& $
     [0,0,0]_{(0,0)}
     $ \\ \hline
     \end{tabular}}
 \caption{Long Konishi multiplet ${\cal
A}^{2}_{[0,0,0](0,0)}$}\label{Kon13}
\end{table}


\begin{thebibliography}{99}

\bibitem{polwl} 
A.~M.~Polyakov,
Nucl.\ Phys.\ B {\bf 120} (1977) 429.
Nucl.\ Phys.\ Proc.\ Suppl.\  {\bf 68} (1998) 1
[hep-th/9711002].

\bibitem{thooN}
G.~'t Hooft,
Nucl.\ Phys.\ B {\bf 75} (1974) 461. 
[hep-th/0204069].

\bibitem{jm}
J.~M.~Maldacena,
Adv.\ Theor.\ Math.\ Phys.\  {\bf 2}, 231 (1998)
[Int.\ J.\ Theor.\ Phys.\  {\bf 38}, 1113 (1999)]
[hep-th/9711200].


\bibitem{adsquant}
R.~Kallosh and A.~A.~Tseytlin,
JHEP {\bf 9810} (1998) 016
[hep-th/9808088].
R.~R.~Metsaev and A.~A.~Tseytlin,
Nucl.\ Phys.\ B {\bf 533} (1998) 109
[hep-th/9805028].
N.~Drukker, D.~J.~Gross and A.~A.~Tseytlin,
JHEP {\bf 0004} (2000) 021
[hep-th/0001204].

\bibitem{berko}
N.~Berkovits, C.~Vafa and E.~Witten,
JHEP {\bf 9903} (1999) 018
[hep-th/9902098].
N.~Berkovits,
JHEP {\bf 0204} (2002) 037
[hep-th/0203248].
N.~Berkovits and O.~Chandia,
Nucl.\ Phys.\ B {\bf 596} (2001) 185
[hep-th/0009168].
N.~Berkovits,
Int.\ J.\ Mod.\ Phys.\ A {\bf 16} (2001) 801
[hep-th/0008145].
N.~Berkovits,
Class.\ Quant.\ Grav.\  {\bf 17} (2000) 971
[hep-th/9910251].


\bibitem{ppwaves}
M.~Blau, J.~Figueroa-O'Farrill, C.~Hull and G.~Papadopoulos,
Class.\ Quant.\ Grav.\  {\bf 19}, L87 (2002)
[hep-th/0201081];
JHEP {\bf 0201}, 047 (2002)
[hep-th/0110242].
R.~R.~Metsaev and A.~A.~Tseytlin,
Phys.\ Rev.\ D {\bf 65} (2002) 126004
[hep-th/0202109].
J.~G.~Russo and A.~A.~Tseytlin,
JHEP {\bf 0204} (2002) 021
[hep-th/0202179].
R.~R.~Metsaev and A.~A.~Tseytlin,
Russ.\ Phys.\ J.\  {\bf 45} (2002) 719
[Izv.\ Vuz.\ Fiz.\  {\bf 2002N7} (2002) 67].


\bibitem{bmn}
D.~Berenstein, J.~M.~Maldacena and H.~Nastase,
JHEP {\bf 0204} (2002) 013
[hep-th/0202021].

\bibitem{sezsun5} E.~Sezgin and P.~Sundell,
JHEP {\bf 0109} (2001) 036
[hep-th/0105001].
JHEP {\bf 0109} (2001) 025
[hep-th/0107186].
Nucl.\ Phys.\ B {\bf 644} (2002) 303
[hep-th/0205131].

\bibitem{sundb}
B.~Sundborg,
Nucl.\ Phys.\ Proc.\ Suppl.\  {\bf 102} (2001) 113
[hep-th/0103247];
P.~Haggi-Mani and B.~Sundborg,
JHEP {\bf 0004}, 031 (2000)
[hep-th/0002189];
A.~K.~Bengtsson, I.~Bengtsson and L.~Brink,
Nucl.\ Phys.\ B {\bf 227}, 41 (1983);
Nucl.\ Phys.\ B {\bf 227}, 31 (1983).


\bibitem{mikha}
A.~Mikhailov,
hep-th/0201019.

\bibitem{vas4}
S.~E.~Konshtein and M.~A.~Vasiliev,
Nucl.\ Phys.\ B {\bf 312} (1989) 402.
M.~A.~Vasiliev,
Int.\ J.\ Mod.\ Phys.\ D {\bf 5} (1996) 763
[hep-th/9611024].
M.~A.~Vasiliev,
hep-th/9910096.
M.~A.~Vasiliev,
Fortsch.\ Phys.\  {\bf 48} (2000) 223.
M.~A.~Vasiliev,
hep-th/0104246.
M.~A.~Vasiliev,
hep-th/0304049.
M.~A.~Vasiliev and E.~S.~Fradkin,
JETP Lett.\  {\bf 44} (1986) 622
[Pisma Zh.\ Eksp.\ Teor.\ Fiz.\  {\bf 44} (1986) 484].

\bibitem{vas5}
S.~E.~Konstein, M.~A.~Vasiliev and V.~N.~Zaikin,
JHEP {\bf 0012} (2000) 018
[hep-th/0010239].
M.~A.~Vasiliev,
Nucl.\ Phys.\ B {\bf 616} (2001) 106
[Erratum-ibid.\ B {\bf 652} (2003) 407]
[hep-th/0106200].
K.~B.~Alkalaev and M.~A.~Vasiliev,
Nucl.\ Phys.\ B {\bf 655} (2003) 57
[hep-th/0206068].

\bibitem{vasD}
M.~A.~Vasiliev,
hep-th/0304049.

\bibitem{klepol}
I.~R.~Klebanov and A.~M.~Polyakov,
Phys.\ Lett.\ B {\bf 550} (2002) 213
[hep-th/0210114].


\bibitem{sezsun4} E.~Sezgin and P.~Sundell,
JHEP {\bf 0207} (2002) 055
[hep-th/0205132].
J.~Engquist, E.~Sezgin and P.~Sundell,
Class.\ Quant.\ Grav.\  {\bf 19} (2002) 6175
[hep-th/0207101].
hep-th/0211113.

\bibitem{pet} A.C.~Petkou,
JHEP {\bf 0303} (2003) 049
[hep-th/0302063].
R.G.~Leigh and A.C.~Petkou,
hep-th/0304217.
E.~Sezgin, P.~Sundell
hep-th/0305040.

\bibitem{gir} L.~Girardello, M.~Porrati and A.~Zaffaroni,
hep-th/0212181.

\bibitem{dasjev}
S.~R.~Das and A.~Jevicki,
hep-th/0304093.

\bibitem{surya}
N.~V.~Suryanarayana,
hep-th/0304208.



\bibitem{frasag} D.~Francia and A.~Sagnotti,
Phys.\ Lett.\ B {\bf 543} (2002) 303
hep-th/0212185.
P.~de Medeiros and C.~Hull,
hep-th/0303036.

\bibitem{strisol} S.~S.~Gubser, I.~R.~Klebanov and A.~M.~Polyakov,
Nucl.\ Phys.\ B {\bf 636} (2002) 99
[hep-th/0204051]; 
A.~A.~Tseytlin,
Theor.\ Math.\ Phys.\  {\bf 133} (2002) 1376
[Teor.\ Mat.\ Fiz.\  {\bf 133} (2002) 69]
[hep-th/0201112].
S.~Frolov and A.~A.~Tseytlin,
JHEP {\bf 0206} (2002) 007
[hep-th/0204226].
A.~A.~Tseytlin,
Int.\ J.\ Mod.\ Phys.\ A {\bf 18} (2003) 981
[hep-th/0209116].
A.~A.~Tseytlin,
hep-th/0304139.
S.~Frolov and A.~A.~Tseytlin,
hep-th/0304255.
A.~Armoni, J.~L.~Barbon and A.~C.~Petkou,
JHEP {\bf 0206} (2002) 058
A.~Armoni, J.~L.~Barbon and A.~C.~Petkou,
JHEP {\bf 0210} (2002) 069
[hep-th/0209224].

\bibitem{konan} 
D.~Amati, K.~Konishi, Y.~Meurice, G.~C.~Rossi and G.~Veneziano,
Phys.\ Rept.\  {\bf 162}, 169 (1988).
D.~Anselmi, J.~Erlich, D.~Z.~Freedman and A.~A.~Johansen,
Phys.\ Rev.\ D {\bf 57} (1998) 7570
[hep-th/9711035].
D.~Anselmi, D.~Z.~Freedman, M.~T.~Grisaru and A.~A.~Johansen,
Phys.\ Lett.\ B {\bf 394} (1997) 329
[hep-th/9608125]; Nucl.\ Phys.\ B {\bf 526} (1998) 543
[hep-th/9708042].

\bibitem{bkrslog}
M.~Bianchi, S.~Kovacs, G.~Rossi and Y.~S.~Stanev,
JHEP {\bf 9908} (1999) 020
[hep-th/9906188].


\bibitem{bkrsandim}
M.~Bianchi, S.~Kovacs, G.~Rossi and Y.~S.~Stanev,
Nucl.\ Phys.\ B {\bf 584} (2000) 216
[hep-th/0003203].


\bibitem{bkrskon}
M.~Bianchi, S.~Kovacs, G.~Rossi and Y.~S.~Stanev,
JHEP {\bf 0105} (2001) 042
[hep-th/0104016].


\bibitem{bersmix}
M.~Bianchi, B.~Eden, G.~Rossi and Y.~S.~Stanev,
Nucl.\ Phys.\ B {\bf 646} (2002) 69
[hep-th/0205321].

\bibitem{appss} 
G.~Arutyunov, S.~Penati, A.~C.~Petkou, A.~Santambrogio and E.~Sokatchev,
Nucl.\ Phys.\ B {\bf 643} (2002) 49
[hep-th/0206020].

\bibitem{dolosb}
F.~A.~Dolan and H.~Osborn,
hep-th/0209056.


\bibitem{qcdetal} 
A.~V.~Kotikov and L.~N.~Lipatov,
hep-ph/0112346;
hep-ph/0208220.
M.~Axenides, E.~Floratos and A.~Kehagias,
hep-th/0210091.
A.~V.~Kotikov, L.~N.~Lipatov and V.~N.~Velizhanin,
Phys.\ Lett.\ B {\bf 557} (2003) 114
[hep-ph/0301021].


\bibitem{kk}
B.~Biran, A.~Casher, F.~Englert, M.~Rooman and P.~Spindel,
Phys.\ Lett.\ B {\bf 134}, 179 (1984).
A.~Casher, F.~Englert, H.~Nicolai and M.~Rooman,
Nucl.\ Phys.\ B {\bf 243}, 173 (1984).
S.~Deger, A.~Kaya, E.~Sezgin and P.~Sundell,
Nucl.\ Phys.\ B {\bf 536}, 110 (1998)
[hep-th/9804166].

\bibitem{deboer} 
A.~Salam and J.~Strathdee,
Annals Phys.\  {\bf 141}, 316 (1982).
J. de Boer,
Nucl. Phys. {\bf B548} (1999) 139, hep-th/9806104;
JHEP {\bf 9905} (1999) 017, hep-th/9812240.
E. Gava, A.B. Hammou, J.F. Morales and K.S.Narain,
JHEP {\bf 0103} (2001)035, hep-th/0102043; hep-th/0201265.


\bibitem{ms}
J.~F.~Morales and H.~Samtleben,
JHEP {\bf 0208} (2002) 042, hep-th/0206247;
hep-th/0211278.


\bibitem{bits}
A.~Karch,
hep-th/0212041.
A.~Clark, A.~Karch, P.~Kovtun and D.~Yamada,
hep-th/0304107.
A.~Dhar, G.~Mandal and S.~R.~Wadia,
hep-th/0304062.
C.~B.~Thorn,
Phys.\ Rev.\ D {\bf 56}, 6619 (1997)
[hep-th/9707048].
H.~Verlinde,
hep-th/0206059.
J.~G.~Zhou,
Phys.\ Rev.\ D {\bf 67}, 026010 (2003)
[hep-th/0208232].
D.~Vaman and H.~Verlinde,
hep-th/0209215.
S.~Bellucci and C.~Sochichiu,
hep-th/0302104.

\bibitem{lind}
A.~Karlhede and U.~Lindstrom,
Class.\ Quant.\ Grav.\  {\bf 3}, L73 (1986);
H.~Gustafsson, U.~Lindstrom, P.~Saltsidis, B.~Sundborg and R.~von Unge,
        Nucl.\ Phys.\ B {\bf 440}, 495 (1995)
       [hep-th/9410143].

\bibitem{lindsun}
U. Lindstrom, B. Sundborg, G. Theodoridis.
Phys.\ Lett.\ B {\bf 253}, 319 (1991); 
Phys.\ Lett.\ B {\bf 258}, 331 (1991).

\bibitem{zeroten}
G.~K.~Savvidy,
Phys.\ Lett.\ B {\bf 552}, 72 (2003),
hep-th/0304160.

\bibitem{lindzab}
U.~Lindstrom and M.~Zabzine,
hep-th/0305098.

\bibitem{ans} D.~Anselmi,
Nucl.\ Phys.\ B {\bf 541} (1999) 369
[hep-th/9809192].

\bibitem{bks}
N.~Beisert, C.~Kristjansen and M.~Staudacher,
hep-th/0303060.

\bibitem{metsa}
R.~R.~Metsaev,
Nucl.\ Phys.\ B {\bf 563}, 295 (1999)
[hep-th/9906217].
Phys.\ Lett.\ B {\bf 468}, 65 (1999)
[hep-th/9908114].
hep-th/9911016.
hep-th/0002008.
Int.\ J.\ Mod.\ Phys.\ A {\bf 16S1C}, 994 (2001)
[hep-th/0011112].
L.~Brink, R.~R.~Metsaev and M.~A.~Vasiliev,
Nucl.\ Phys.\ B {\bf 586}, 183 (2000)
[hep-th/0005136].
R.~R.~Metsaev, C.~B.~Thorn and A.~A.~Tseytlin,
Nucl.\ Phys.\ B {\bf 596}, 151 (2001)
[hep-th/0009171].

\bibitem{polya} G.~Polya and R.~C.~Read, ``Combinatorial Enumeration 
of Groups, Graphs and Chemical Compounds'', Springer Verlag, New York 
1987.

\bibitem{polpol}
A.~M.~Polyakov,
Int.\ J.\ Mod.\ Phys.\ A {\bf 17S1} (2002) 119
[hep-th/0110196].

\bibitem{gsw} M.~B.~Green, J.~H.~Schwarz and E.~Witten,
``Superstring Theory'', Cambridge University Press 1987.

\bibitem{andfer}
L.~Andrianopoli and S.~Ferrara,
Lett.\ Math.\ Phys.\  {\bf 48} (1999) 145 [hep-th/9812067].


\bibitem{ffz}
S.~Ferrara, C.~Fronsdal and A.~Zaffaroni,
Nucl.\ Phys.\ B {\bf 532} (1998) 153
[hep-th/9802203].
S.~Ferrara and A.~Zaffaroni,
hep-th/9908163.

\bibitem{gm}
M.~Gunaydin and N.~Marcus,
Class.\ Quant.\ Grav.\  {\bf 2} (1985) L11.

\bibitem{aeps} 
G.~Arutyunov, B.~Eden, A.~C.~Petkou and E.~Sokatchev,
Nucl.\ Phys.\ B {\bf 620} (2002) 380
[hep-th/0103230];
G.~Arutyunov, B.~Eden and E.~Sokatchev,
Nucl.\ Phys.\ B {\bf 619} (2001) 359
[hep-th/0105254].
B.~Eden and E.~Sokatchev,
Nucl.\ Phys.\ B {\bf 618} (2001) 259
[hep-th/0106249].

\bibitem{dhhhr}
E.~D'Hoker, P.~Heslop, P.~Howe and A.~V.~Ryzhov,
hep-th/0301104.


\bibitem{wittetal} F.~Cachazo, M.~R.~Douglas, N.~Seiberg and E.~Witten,
JHEP {\bf 0212} (2002) 071
[hep-th/0211170].

\bibitem{prep} M.~Bianchi, B.~Eden, G.~Rossi and Y.~S.~Stanev,
to appear.

\bibitem{kutlar} D.~Kutasov and F.~Larsen,
JHEP {\bf 0101} (2001) 001
[hep-th/0009244].

\bibitem{witjhs} E. Witten, Talk at John Schwarz 60-th Birthday 
Symposium,\\
http://theory.caltech.edu/jhs60/witten/1.html

\bibitem{beis}
N.~Beisert,
hep-th/0211032.


\bibitem{bkps}
N.~Beisert, C.~Kristjansen, J.~Plefka and M.~Staudacher,
Phys.\ Lett.\ B {\bf 558} (2003) 229
[hep-th/0212269].

\bibitem{mz} J.~A.~Minahan and K.~Zarembo,
JHEP {\bf 0303} (2003) 013
[hep-th/0212208].

\bibitem{afsz}
L.~Andrianopoli, S.~Ferrara, E.~Sokatchev and B.~Zupnik,
Adv.\ Theor.\ Math.\ Phys.\  {\bf 3} (1999) 1149
[hep-th/9912007].


\bibitem{dp}
V.~K.~Dobrev and V.~B.~Petkova,
Phys.\ Lett.\ B {\bf 162} (1985) 127.




\end{thebibliography}
\end{document}